\documentclass[pre,twocolumn]{revtex4-1}

\usepackage{graphicx}% Include figure files
\usepackage{dcolumn}% Align table columns on decimal point
\usepackage{bm}% bold math
\usepackage[utf8]{inputenc}
\usepackage[T1]{fontenc}
\usepackage{mathptmx}

\usepackage{pst-node}

\usepackage{ulem}
\usepackage{mhchem}
\usepackage{wasysym}
\usepackage{tikz}
\usetikzlibrary{decorations.pathmorphing,patterns}
\usepackage{pst-coil}
\usepackage{lipsum} 
\usetikzlibrary{graphs}

\usepackage{listings}
\usepackage{color}
\definecolor{dkgreen}{rgb}{0,0.6,0}
\definecolor{gray}{rgb}{0.5,0.5,0.5}
\definecolor{mauve}{rgb}{0.58,0,0.82}

\usepackage{verbatim}   % useful for program listings
\usepackage{color}      % use if color is used in text
\usepackage{subfigure}  % use for side-by-side figures
\usepackage{hyperref}   % use for hypertext links, including those to external documents and URLs

\usepackage{braket}

\newcommand{\bnabla}{\mbox{\boldmath $\nabla$}}

\newcommand{\ba}{\begin{eqnarray}}
\newcommand{\ea}{\end{eqnarray}}
\newcommand{\be}{\begin{equation}}
\newcommand{\ee}{\end{equation}}

\begin{document}

%==============================================================
\title{Integral equation formulation of run-and-tumble particles in a harmonic trap:  
the special status of a system in two-dimensions}
%{Interpretation of run-and-tumble motion as jump-process:  the case of a harmonic trap}
%==============================================================

\author{Derek Frydel}
\affiliation{Department of Chemistry, Universidad Técnica Federico Santa María, Campus San Joaquin, Santiago, Chile}

\date{\today}

\begin{abstract}

Statistical-mechanical models often exhibit a dimension-dependent solvability:
in 1D, exact solutions are straightforward; in 2D, solutions are exact but require nontrivial derivations; and in 3D, closed-form solutions are typically unavailable.
This logic is repeated for a simple model of self-propelled particles, run-and-tumble particles (RTP) in a harmonic trap, confirming the claim that the system in 2D enjoys special status.  
This study revisits the RTP-harmonic-trap model using an integral-equation formulation recently proposed in Ref. \cite{POF-Frydel-2024}.  
The formulation is based on reinterpreting RTP motion as a jump process.  
The key quantity of the formulation is a transition operator $G(x,x')$, representing the probability distribution of the jumps of an auxiliary system.  
The stationary distribution is then obtained from the integral equation $\rho(x) = \int dx' \, \rho(x') G(x,x')$.  In 2D, we find that $G(x,x')$ is 
reversible.  This implies the $\rho(x)$ satisfied the detailed balance condition, $\rho(x') G(x,x') = \rho(x) G(x',x)$, from which $\rho(x)$ can be 
obtained without need of an integral equation.  The reversibility of $G(x,x')$ does not mean that RTP particles are in equilibrium.  
It only means that our specific interpretation of the RTP motion leads to an auxiliary system that is in equilibrium. 
%The special status of a system in 2D is found to be a consequence of $G(x,x')$ being reversible, 
%which implies that $\rho(x)$ satisfies the detailed balance relation, $\rho(x') G(x,x') = \rho(x) G(x',x)$, which can used to calculate $\rho(x)$.  
%The special status of a system in 2D is lost if the distribution of waiting times that determines the duration of the "run" stage is a non-exponential 
%function.  
The reversibility of the system in 2D is lost if the probability distribution of the waiting times (the times that determine the duration on the "run" stage of the RTP motion) 
deviates from an exponential distribution.  

\end{abstract}

\pacs{
}

\maketitle
%------------------------------------------------

\section{Introduction}

The distinguishing feature of self-propelled motion is its continuous use of energy, which is 
dissipated into the environment without being recovered.  This leads to the breakdown of the 
fluctuation-dissipation relation and the detailed-balance condition, resulting in non-equilibrium statistical 
mechanics.  Run-and-tumble particle model (RTP) is the simplest and most ideal representation of 
self-propelled motion 
\cite{PRL-Cates-2008,EPL-Cates-2009,COM-Grognot-2021,Malakar-2018,PRE-Razin-2020}.  
It was originally conceived to represent the motion of bacteria \cite{berg-1983,PRE-Schnitzer-1993}.  
Other popular models of self-propelled particles, not considered in this work, are the active Brownian 
particle model 
\cite{EPJ-Ebeling-2012,PRL-Marchetti-2012,RMP-Marchetti-2013,SR-Maggi-2015,EPJ-Cates-2015,PhysA-Fodor-2018,PRE-Shankar-2018,NJP-Gompper-2018,PRL-Caraglio-2022,PRE-Manoj-2023,PRL-Suma-2023b,EPL-Suma-2024,SF-Suma-2024}, more appropriate for representing active colloids \cite{AdvSci-2023}, and 
the active Ornstein-Uhlenbeck particle model \cite{PRE-Szamel-2014,PRE-Fodor-2021,JSTAT-Caprini-2019,PRL-Suma-2023}, 
originally conceived to represent passive particles in an active bath.  Alternatively, an active motion can 
be induced when passive particles are driven by fluctuating external potential 
\cite{PRE-Pal-2013,JOP-Santra-2021,JOPA-Gupta-2020,JSTAT-Gupta-2021,JOPA-Alston-2022,PRE-Schehr-2024,PRE-Frydel-2024}.

RTP motion consists of two alternating stages.  During the deterministic "run" stage, a particle moves at constant
swimming velocity in a fixed orientation, and during the "tumble" stage, which occurs 
instantaneously, a particle changes its orientation.  In the standard RTP model, the duration of the "run" stage 
is drawn from an exponential distribution $p_t$, which corresponds to a constant tumbling rate.    
%Any other functional form of $p_t$ implies a non-constant rate 
%of tumbling, which could be represented by a position-dependent rate \cite{PRE-Farago-2024}.  

Various aspects of the RTP model have been investigated in the past.  
The RTP model in one-dimension has been extensively analyzed in \cite{Malakar-2018}.  
The entropy production rate %of the one-dimensional system 
was considered in \cite{PRE-Razin-2020,PRE-Fyrdel-2022,PRE-Gupta-2023}.  
Dynamic properties, including first passage, survival probability, and local time were 
studied in \cite{EPJE-Angelani-2014,PRL-Schehr-2020,JSTAT-Scher-2021,PRE-Kundu-2021}.  
Extensions of the RTP model in one-dimension to include
more than two discrete swimming orientations were considered in \cite{JPA-Basu-2020,JSTAT-Frydel-2021,POF-Frydel-2022,EPJE-Hartmut-2022}.  
A harmonic potential in one-dimension was considered in 
\cite{PRL-Cates-2008,EPL-Cates-2009,JPA-Basu-2020,PRE-106-Frydel-2022,PRE-Scher-2022,Smith-PRE-2022a,POF-Frydel-2023}.  
RTP particles in other types of potentials and under different conditions has been considered in 
\cite{PRE-Dhar-2019,PRE-Farago-2024,PRE-connor-2023,PRE-Dhar-2019,EPL-Doussal-2020,JSTAT-Wijland-2023,JPA-Angelani-2015,JSTAT-Bressloff-2023,PRL-Tailleur-2019,EPL-Detcheverry-2015,PRE-Dean-2021,JPA-Angelani-2017}.

In this work, we consider RTP particles in a harmonic trap under steady-state conditions within the integral-equation formulation 
recently introduced for active systems in confinement between two walls \cite{POF-Frydel-2024}.  The re-formulation in that 
reference established that the system of RTP particles in a stationary state and a splitting probability problem share the same 
theoretical framework \cite{PRL-Klinger-2022}.  This analogy led to new insights and some exact analytical results.  In this work, 
we want to apply the integral equation formulation to RTP particles in a harmonic trap to gain a deeper theoretical understanding.  

Our specific focus is on the system in 2D.  The solution in 1D has been known for some time and, to our knowledge,
was first published in \cite{PRL-Cates-2008}.  The solution in this case is relatively straightforward as the orientation 
of swimming velocities is limited to two discrete values, $v = \pm v_0$.  In a more recent work \cite{PRE-106-Frydel-2022}, the solution 
in 2D was obtained by analyzing moments of a stationary distribution.  The sequence of moments was then identified with a specific distribution.  
Although correct, this type of derivation is more inferential than rigorous, and one of the motivations to explore an alternative formulation
of the problem is to seek a more rigorous derivation and provide a deeper understanding of the system in 2D.  For example, why the system in 2D
admits of an exact solution, but no such solution is available in 3D.  

In the standard RTP motion, 
the duration of the "run" stage is a random variable drawn from an exponential distribution $p_t = \tau^{-1} e^{-t/\tau}$.  Since the exponential $p_t$
corresponds to a constant tubling rate and represents a memoryless process, 
the second aim of this work is to determine the role of non-exponential distributions of the waiting-times. In this part of the work 
the integral equation formulation is extended to non-exponential $p_t$.   
%In that part, we connect to recent work \cite{PRE-Farago-2024} which studied RTP motion in 1D with 
%non-exponential waiting-time distributions.

The general organization of this paper is as follows.  
In Sec. (\ref{sec:sec1}) we introduce the model.  In Sec. (\ref{sec:sec3}), we derive the integral formulation 
for RTP particles in a harmonic trap.   In Sec. (\ref{sec:sec2}) we extend the integral equation 
framework to treat the model with non-exponential distribution of waiting times.  We conclude the work with
Sec. (\ref{sec:sec4}).

\section{The model}
\label{sec:sec1}

RTP motion consists of two alternating stages.  During the "run" stage, which is deterministic, 
a particle moves at a constant swimming velocity $v_0$ in a direction designated by the unit 
vector ${\bf u}$.  The duration of the "run" stage is determined by the persistence time $t_w$ 
drawn from the exponential distribution $p_t \propto e^{-t_w/\tau}$, where $\tau$ represents 
the average persistence time (exponential $p_t$ indicates that tumbling events occur at a 
constant rate).  During the "tumble" stage, which occurs instantaneously, a particle 
changes its orientation to any other orientation, ${\bf u} \to {\bf u}'$.  
If a particle is trapped in an external potential $u({\bf r})$, there is an additional force 
${\bf F} = -\bnabla u$ acting on a particle.    

Within the Fokker-Planck formulation, the probability distribution of RTP particles (at zero temperature 
and for convenience assumed to occur in two-dimensions) evolves as \cite{pressure-2015,JSTAT-Frydel-2021}
\be
\frac{\partial n}{\partial t} = -\bnabla\cdot \left[ \left({\bf F}  +  v_0 {\bf u} \right) n \right]  
 -    \frac{ 1 }{\tau} \left(  n  -   \frac{1}{2\pi} \int_{0}^{2\pi} d\theta\, n({\bf r},\theta,t)\right),
\label{eq:FP2D}
\ee
where $n\equiv n({\bf r},\theta,t)$ is the probability distribution.  The first term on the right-hand side 
is connected to the flux, and the second term 
governs the evolution of a unit vector ${\bf u} = (\cos\theta,\sin\theta)$ representing the swimming orientation
in two-dimensions whereby it gives rise to active motion. 
(In Appendix (\ref{sec:app0}) we write down Fokker-Planck equations for dimensions $d=1,3$.)  

In this work, we consider RTP particles confined to a harmonic potential in one direction, 
$u(x) = \frac{K}{2} x^2$.  And because a stationary distribution of such a system is non-uniform 
along the $x$-axis only, the system is effectively one-dimensional.  The projected swimming velocity 
on the $x$-axis, $v = v_0\cos\theta$, results in a probability distribution $p_v(v)$ whose distribution 
within the interval $v\in [-v_0,v_0]$ depends on a system dimension.  

The Fokker-Planck equation of such a system can be represented as 
\be
0   =   \left(\mu K x  -  v\right) n'   -    \frac{ 1 }{\tau} \left(  n  -   \int_{-v_0}^{v_0} dv\, p_v n\right),
\label{eq:FP}
\ee
where $n\equiv n(x,v)$ is the time-independent density distribution.  Note that by transforming Eq. (\ref{eq:FP2D})
to Eq. (\ref{eq:FP}), an integral over $\theta$ transforms to an integral over $v$ as a result 
of a change of a variable $v = v_0 \cos\theta$.  The advantage of the formulation in Eq. (\ref{eq:FP})
is that it applies to all dimensions.  The dependence on a system dimension enters via 
$p_v(v)$, 
\be
p_v
=  \frac{1}{2}
\begin{cases}
     \delta(v-v_0) + \delta(v+v_0), & \text{for $d=1$},\\
     \frac{2}{\pi} \frac{1}{\sqrt{v_0^2-v^2}}, & \text{for $d=2$}, \\
      \frac{1}{v_0}, & \text{for $d=3$}. \\
  \end{cases}
\label{eq:pv}
\ee
For $d=1$ there are only two possible orientations of motion, $v=\pm v_0$, and so the distribution
is represented by two delta functions \cite{PRE-Razin-2020}.  The distributions for $d=2$ and 
$d=3$ are obtained using the change of a variable $v = v_0 \cos\theta$ that modifies the integral 
over an angular orientation as
$$
 \int_0^{2\pi} d\theta  ~\to~  
= \int_{-v_0}^{v_0} dv\,\frac{1}{\sqrt{v_0^2-v^2}} 
$$
for $d=2$, and 
$$
 \int_0^{2\pi} d\phi \int_0^{\pi} d\theta\, \sin\theta   
~\to~  
 \int_{-v_0}^{v_0} dv
$$
for $d=3$.  A marginal stationary distribution is defined as 
\be
\rho(x) = \int_{-v_0}^{v_0} dv\, p_v(v) n(x,v).  
\ee
In Appendix (\ref{sec:app0a}), we provide more details of the two transformations.  

In this work, we re-formulate the problem of RTP particles in a stationary state as an integral equation 
involving $\rho(x)$ rather than $n(x,v)$ as it appears in the FP formulation in Eq. (\ref{eq:FP}).  
This means that we will not analyze Eq. (\ref{eq:FP}).  Instead, we will analyze an alternative formulation 
of the same system.  
We obtain such a re-formulation by re-interpreting the RTP motion as a jump-process, the details of which we provide in the next section.

\section{harmonic potential within jump-process interpretation}
\label{sec:sec3}

One of the aims of this work is to study RTP particles in a harmonic trap under stationary conditions 
by considering the integral equation formulation, which is possible by re-interpretting the RTP motion as a jump process.  
To arrive at such a framework, we use a technique recently proposed and applied to study RTP 
particles confined between two walls \cite{POF-Frydel-2024}.  
%The re-formulation of the problem led to some exact results and 
%established an analogy between RTP particles between the walls and the splitting probability problem \cite{PRL-Klinger-2022}.  

The motivation to apply this framework to harmonic confinement is to gain a deeper theoretical understanding of this system.  
%The stationary distribution in dimension $d=1$ is known since
%at least it was  \cite{PRL-Cates-2008}.  The stationary distribution in two dimensions was obtained indirectly by analyzing 
%moments of a stationary distribution \cite{PRE-106-Frydel-2022}.  From those moments, it was possible to infer a functional form of the 
%distribution; thus, the distribution was inferred rather than derived.  By considering the same problem in an alternative framework, 
%we hope to provide a rigorous derivation.  Finally, for the case in three-dimensions, at the moment, no exact analytical expression
%for a stationary distribution is known.  It is hoped that analyzing this case in an alternative framework can provide 
%additional insights and at least a numerical scheme to calculate the distribution as an alternative to simulations. 
We are specifically interested in the two-dimensional case, as this case admits an exact solution, and we want to 
better understand what precise conditions make the exact solution possible.  
%We want to better understand why, in this dimension 
%it is possible to get an exact solution.  The two-dimensional systems often enjoy the status of being exactly solvable.  

%The things stand as follows.  The system in one-dimension is often trivial, then the case of two dimensions admits 
%exact solution after some time and creativity has been put into the effort, and the case in three-dimenisions 
%is 
%
%many statistical mechanics models can be solved exactly in 2D, but the leap to 3D encounters major mathematical 
%and computational roadblocks.  The fact that it is possible to obtain an exact solution for RTP particles in a harmonic 
%trap but it is impossible to obtain in three- and higher dimensions, confirms this trend.  As the solution for the case
%in 2D was obtained more by inference, deduced from moments of the distributions that can be calculated 
%exactly, the more theoretical reason or context feels to be missing.  Motivated by a recent work on RTP particles 
%between two walls, 

The integral equation formulation stems from the interpretation of the RTP motion as a jump process, where a 
particle, rather than moving continuously in time, jumps between positions corresponding to the "tumble" stage.  
%of the RTP motion, and ignoring configurations of the "run" stage, the obtained distributions are correct.  
%algorithm that neglects configurations generated during the "run" phase, 
%and only configurations of the "tumble" phase are retained. 
%As a result of this biased sampling method, a particle appears to jump between positions rather than 
%move continuously in space and time—hence, we refer to this approach as a jump-process interpretation.  
The key quantity of this formulation is the probability distribution of particle jumps along the $x$-axis for a particle
initially at $x'$.  If we designate this distribution by  $G(x,x')$, then the stationary distribution of particles is 
obtained from the following integral equation 
\be
\rho_{}(x)  =  \int dx'\, \rho_{}(x') G(x,x'). 
\label{eq:integral}
\ee
The relation above tells us that the distribution does not change even after a particle
jumps to another position drawn from the distribution $G(x,x')$.  

We would like to point out the similarity of Eq. (\ref{eq:integral})
to a similar integral equation in \cite{POF-Frydel-2024} for RTP particles between two walls.  There are, however, 
some notable differences.  For a slit geometry, the distirbution of jumps depends only on the distance $|x-x'|$, 
$G(x,x') = G(|x-x'|)$.  In the equation in Ref. \cite{POF-Frydel-2024}, there are additional terms due to a fraction of particles adsorbed onto 
the walls.

The integral equation of Eq. (\ref{eq:integral}) is valid for any general external potential, and the main challenge, before the equation can be used, is to obtain $G(x,x')$.  
To calculate $G(x,x')$, we note that the motion during the "run" stage is governed by the deterministic equation
\be
\dot x =  -\mu K x + v, 
\label{eq:langevin}
\ee
for which the solution is 
\be
x(t)    =    x_0 e^{-\mu K t}    +    \left( 1  -  e^{-\mu K t} \right)  \frac{v}{\mu K},
\label{eq:xtp}
\ee
where $x_0$ is the position at time $t=0$.  
Because the motion is deterministic, the probability distribution of a particle at a given time is a propagating delta function.  
Because the position of a particle at the end of the "run" stage depends on 
the swimming velocity $v$ and the waiting time $t$, both random variables drawn from 
$p_t$ and $p_v$, respectively, the distribution of jumps $G(x,x_0)$ corresponds to the averaged delta propageting distribution
given by 
\be
G(x,x_0) = \int_0^{\infty} dt \, p_t \int_{-v_0}^{v_0} dv \, p_v  \delta\left( x -  x_0  e^{-\mu K t}     
-    \frac{v\left( 1  -  e^{-\mu K t} \right)}{\mu K}  \right).  
\label{eq:G-00}
\ee
The delta function of the integrand is the distribution of a deterministic motion in Eq. (\ref{eq:xtp}).

Integrating Eq. (\ref{eq:G-00}) over time yields an expression that depends on the relative position of $x$ with respect to $x_0$,   
\ba
&&G_+(x,x_0<x)%(x,x_0<x) 
= \alpha \int_{x}^{x_m} dy\,  \frac{p(y)}{y - x_0}  \left( \frac{y - x}{y - x_0} \right)^{\alpha-1} \nonumber\\
&&G_-(x,x_0>x)%(x,x_0>x) 
= \alpha  \int_{-x_m}^{x} dy\,   \frac{p(y)}{x_0 - y}  \left( \frac{y - x}{y - x_0} \right)^{\alpha-1}, 
\label{eq:Gpm}
\ea
where 
$
\alpha = \frac{1}{\tau \mu K},
$
is the dimensionless tumbling rate, 
$
x_m = \frac{v_0}{\mu K}
$
is the maximal distance from the trap center that a particle can reach, and 
$p(y)$ is related to $p_v$ in Eq. (\ref{eq:pv}) and depends on the system dimension, 
\be
p(y)
=  \frac{1}{2}
\begin{cases}
     \delta(x_m-y) + \delta(x_m+y), & \text{for $d=1$},\\
     \frac{2}{\pi} \frac{1}{\sqrt{x_m^2-y^2}}, & \text{for $d=2$}, \\
      \frac{1}{x_m}, & \text{for $d=3$}. \\
  \end{cases}
\label{eq:py}
\ee
See Appendix (\ref{sec:app0c}) for details of how Eq. (\ref{eq:Gpm}) is derived.

The evaluated form of Eq. (\ref{eq:Gpm}) depends on the system dimension $d$.  Regardless of $d$, 
however, $G(x,x')$ always vanishes at $x=\pm x_m$, unless $x_0=\pm x_m$.  For $d>1$, $G(x,x')$ 
exhibits a logarithmic singularity at $x=x'$.  Such a logarithmic divergence has already been observed 
in $G(x,x')$ for RTP particles between two walls \cite{POF-Frydel-2024}.  
Finally, for $d=1$ and $d=3$ the operator $G(x,x')$ is discontinuous at $x=x_0$.  If we define
the discontinuity function as 
\be
\Delta G = \lim_{h\to 0} \big[ G(x+h,x) - G(x - h,x) \big]
\label{eq:DG-def}
\ee 
then the discontinuity at $x=x'$ for different $d$ is given by 
\be
 \Delta G(x)
=  
\begin{cases}
      \frac{\alpha x}{x_m^2 - x^2} , & \text{for $d=1$},\\
      0, & \text{for $d=2$}, \\
       \frac{\alpha}{2 x_m} \ln \big( \frac{x_m - x }{x_m + x} \big) , & \text{for $d=3$}, \\
  \end{cases}
\label{eq:DG}
\ee
%where 
%\be
%\Delta G = \lim_{h\to 0} \big[ G(x+h,x) - G(x - h,x) \big]
%\label{eq:DG-def}
%\ee 
Alternatively, $\Delta G(x)$ can be defined as $\Delta G = \lim_{h\to 0} \big[ G(x+h,x) - G(x,x+h) \big]$.

We have managed to map the RTP model in a harmonic trap into an auxiliary system governed by the 
transition operator \( G(x, x') \), which encapsulates all the relevant information of the original system, such as 
an external potential and the underlying dynamics.

Up until now, we have restricted our discussion to mathematical features of \( G(x, x') \) and said nothing 
about its physical significance.  To begin with, \( G(x, x') \) can be interpreted as the transition probability between 
two arbitrary points \( x \) and \( x' \).  We could, for example, try to determine whether  \( G(x, x') \) is reversible.  
Even if the original system of RTP particles out-of-equilibrium, we cannot exclude the possibility that the auxiliary system is in equilibrium.

The Kolmogorov loop criterion permits us to determine whether  \( G(x, x') \) is reversible over 
a closed loop of states \cite{Glynn-2009}. The simplest loop is the traingular loop involving three states \( x \), \( x' \), and \( x'' \), 
and the operator \( G(x, x') \) is reversible if 
\begin{equation}
G(x, x') G(x', x'') G(x'', x) = G(x, x'') G(x'', x') G(x', x). 
\label{eq:KC1}
\end{equation}
This condition implies that the system is reversible and that the stationary distribution \( \rho(x) \) satisfies detailed balance~\cite{PRE-Ramon-2024}:
\be
\rho(x')G(x,x')=\rho(x)G(x',x).  
\label{eq:DBC}
\ee
Otherwise, if Eq.~(\ref{eq:KC1}) is not satisfied, the system is irreversible.   

Without considering complete expressions of $G(x,x')$, we can determine whether the discontinuity  
$\Delta G(x)$ may play a role in $G(x,x')$ being reversible or not.  
%For example, we may want to know if the presence of the discontinuity in \( G(x, x') \) at $x=x'$ 
%for system dimensions $d=1$ and $d=3$, see Eq. (\ref{eq:DG}), has physical siginficance beyond mathematics and somehow
%it affects a system's reversibility or its absence.      
If we define $x'=x+h$, then take the limit $h\to 0$ and assume that $\rho(x)$ is continuous in the region $x\in(-x_m,x_m)$, 
then the detailed balance condition in Eq. (\ref{eq:DBC}) is found to 
yield $G(x,x+h)=G(x+h,x)$.  However, if we recall Eq. (\ref{eq:DG}), then we can identify this quantity with the discontinuity at $x=x'$, 
$G(x+h,x) - G(x,x+h) = \Delta G(x)$.  Thus, for $\Delta G \neq 0$, the operator $G(x,x')$ is irreversible.  
Only for the dimension $d=2$, where $\Delta G = 0$, the operator $G(x,x')$ can be reversible.  We do not claim that the absence
of discontinuity is a sufficient condition of reversibility; however, we can establish that it is a necessary condition.

%We conclude that the presence of a discontinuity in \( G(x, x') \) at $x=x'$ for $d=1$ and $d=3$ excludes the possibility of 
%an auxiliary system to be reversible.  It leaves the system in $d=2$ as the only case which may permit reversibility.
%As we will see in subsequent sections, it turns out that the auxiliary system for $d=2$ is indeed reversible.  This 
%significantly reduces the mathematical complexity of the theoretical framework and leads to a simple analytical solution.  

%\subsection{simulation algorithm}

To provide a more intuitive understanding of the jump-process interpretation of the RTP motion in a harmonic trap, 
we present a simulation algorithm based on this interpretation.  
The particle's motion within the jump-process interpretation is characterized by discrete jumps governed by 
\be
x_{n+1} = x_n e^{-\mu K t_w}  +   \left( 1 -  e^{-\mu K t_w} \right) \frac{v}{\mu K},
\label{eq:xn}
\ee
where we used Eq. (\ref{eq:xtp}) to determine the position of a particle at the next step.  
Note that a new position depends on two random variables, the waiting time $t_w$, which is drawn from an exponential
distribution, and $v$ drawn from $p_v$, which depends on a system dimension and is given in Eq. (\ref{eq:pv}).

\subsection{case $d=1$}
\label{sec:sec3b}

Because for $d=1$ there are only two possible directions of motion, $v_{}= \pm v_0$, a stationary 
distribution $\rho$ can be obtained directly from the Fokker-Planck formulation 
\cite{PRL-Cates-2008,EPL-Cates-2009,PRE-106-Frydel-2022}, which consists of two coupled differential 
equations that can be solved exactly --- see Eq. (\ref{eq:app0-4}) in Appendix (\ref{sec:app0}).

We revisit the system in $d=1$ to confirm the validity of the framework based on the transition operator $G(x,x')$.  
For $d=1$, Eq. (\ref{eq:Gpm}) evaluates to 
\ba
&&G_+(x,x') = \frac{\alpha}{2}      \frac{1}{x_m - x'}   \left(  \frac{ x_m - x } { x_m  - x' }   \right)^{\alpha-1} \nonumber\\
&&G_-(x,x') = \frac{\alpha}{2}      \frac{1}{x_m + x' }  \left(  \frac{ x_m + x } { x_m + x' }  \right)^{\alpha-1}.  
\label{eq:G1D}
\ea
Due to discontinuity of $G(x,x')$ at $x=x'$, we know that $G(x,x')$ is irreversible, and we cannot use the 
detailed balance relation to calculate $\rho(x)$.    
%Thus, both the original system of RTP particles and the auxiliary system are out-of-equilibrium.  

To obtain $\rho(x)$ from $G(x,x')$, we use Eq. (\ref{eq:integral}).  
Due to discontinuity of $G(x,x')$ at $x=x'$, this equation is more conveniently written as 
\be
\rho(x)  =  \int_{-x_m}^{x} dx'\, \rho(x') G_+(x,x') + \int_{x}^{x_m} dx'\, \rho(x') G_-(x,x').  
\label{eq:IE2}
\ee
By differentiating the above equation with respect to $x$, we get the following differential equation 
\be
(x_m^2 - x^2) \rho'  =   x  \rho  -  (\alpha - 1) x_m  q, 
\label{eq:drho}
\ee
where 
\be
q   =   \int_{-x_m}^{x} dx'\, \rho(x') G_+(x,x')   -    \int_{x}^{x_m} dx'\, \rho(x') G_-(x,x').  
\ee
Note that the sole difference between $q$ and $\rho$ in Eq. (\ref{eq:IE2}) is the 
sign in front of the second term.   Taking the derivative of $q$, leads to another differential equation, 
\be
(x_m^2 - x^2) q'   =    x_m \rho  -  (\alpha - 1) x q.  
\label{eq:dq}
\ee
Eq. (\ref{eq:drho}) and Eq. (\ref{eq:dq}) can be combined to yield a second-order differential equation 
$$
0   =     (\alpha - 2) \rho      +     (\alpha - 4)  x \rho'       +      (x_m^2 - x^2)  \rho'', 
$$
which can be reduced to a first-order differential equation given by 
\be
0   =     (\alpha - 2) x\rho        +     (x_m^2   -    x^2) \rho',
\label{eq:diff-1d}
\ee
for which the solution is 
\be
\rho\propto (x_m^2 - x^2)^{\frac{\alpha}{2} - 1}. 
\label{eq:rho1D}
\ee
Details of the derivation can be found in Appendix (\ref{sec:app0d}).

Both Eq. (\ref{eq:diff-1d}) and the solution in Eq. (\ref{eq:rho1D}) agree with the known 
exact results obtained directly from the Fokker-Planck equation.  The re-derivation of those results in this section 
confirms the validity of the alternative formulation of the same problem.

\subsection{case $d=2$}
\label{sec:sec3c}

For the case $d=2$, the formula in Eq. (\ref{eq:Gpm}) evaluates to 
\ba
G(x,x') &=& \frac{ \Gamma(\alpha+1) } { 2^{\alpha} \sqrt{\pi} \Gamma(\alpha+1/2)} \sqrt{\frac{1}{x_m^2 - x'^2}} \left( \frac{x_m^2 - x^2}{x_m^2-x'^2} \right)^{\frac{\alpha}{2} - \frac{1}{2}}
\nonumber\\ 
&&
w^{\alpha/2}   \,\,\,     _2F_1 \left( \frac{\alpha}{2}, \frac{\alpha}{2} + \frac{1}{2}, \alpha + \frac{1}{2}, w   \right),
\label{eq:G2D}
\ea
where $_2F_1$ is the hypergeometric function.  To simplify the nomenclature, we introduce 
a dimensionless parameter 
$$
w = \left[1 + \frac{ (x-x')^2 x_m^2} { (x_m^2-x^2) (x_m^2 - x'^2)} \right]^{-1}.  
$$

Because in Sec. (\ref{sec:sec3}) we determined that $G(x,x')$ for $d=2$ is continuous across $x=x'$, 
we cannot exclude that $G(x,x')$ is reversible.  
%see Eq. (\ref{eq:DG}).  It was also determined that the presence of a discontinuity excludes the possibility 
%of $G(x,x')$ to be reversible.  Its absence, however, does not guarantee that $G(x,x')$ is reversible, and 
%we need additional tests.  
Using the Kolmogorov loop criterion for the 3-state loop given in Eq. (\ref{eq:KC1}), 
we determine that $G(x,x')$ is reversible.  This means that $\rho(x)$ satisfies the detailed balance condition, 
$$
\rho(x) G(x',x) = \rho(x') G(x,x'), 
%\label{eq:DB}
$$
which allows us to define $\rho(x)$ as 
$
\rho(x)  =  \rho(x'){G(x,x')}/{G(x',x)}.
$
For convenience we select $x'=0$, leading to $\rho(x)  =  \rho(0){G(x,0)} / {G(0,x)}$.  Since 
$\rho(0)$ functions as a normalization constant, we ignore it.  
By choosing any other $x'$, we could represent the unnormalized distribution as 
\be
\rho(x)  \propto  \frac{G(x,x')}{G(x',x)}.  
\label{eq:rho-DB}
\ee
Using the formula in Eq. (\ref{eq:G2D}), the stationary distribution based on the above
equation becomes 
\be
\rho(x) \propto \left( x_m^2 - x^2 \right)^{\alpha-\frac{1}{2}},
\label{eq:rho2D}
\ee
whose functional form is the same as that for $\rho$ in Eq. (\ref{eq:rho1D}) --- 
despite the fact that $G(x,x')$ for each case is significantly different.   

%Not only this, but in one case it is irreversible and in another reversible.  

It needs to be emphasized that the original system of RTP particles is not in equilibrium.  
It is only after the system is mapped onto an auxiliary system using jump-process interpretation that it becomes a system in equilibrium.  
Such mapping significantly simplifies mathematical analysis, which leads to an exact and simple solution.  

%The significance of the results of this section is that it provides a rigorous derivation of the stationary 
%distribution using Eq. (\ref{eq:rho-DB}) and Eq. (\ref{eq:G2D}).  It also provides reasons why such derivation was possible, 
%even if those reasons apply to an auxiliary system.  

The fact that the auxiliary system is in equilibrium can be traced to the absence of discontinuity in $G(x,x')$ at $x=x'$.  
%The other consideration to be taken into account is the fact that we used an exponential distribution of the waiting times, 
%$p_t = \tau^{-1} e^{-t / \tau}$.  As an exponential $p_t$ represents a memoryless process, it is reasonable to assume 
%that using any other form of $p_t$ would prevent a system from being reversible.  We will look into other forms of $p_t$ 
%in the second part of the work.  
%Many statistical-mechanics models can be solved exactly in 2D, but the leap to 3D encounters major mathematical 
%challenges.  In other words, there are techniques that apply to a system in 2D but not to 3D.  
%One of the goals of this work is to better understand the special status of the system in 2D within the integral equation framework
%considered in this work.  We determined that the special status is linked to the fact that the transition operator $G(x,x')$ 
%of the auxiliary system is reversible.  We were also able to determine that the reversibility could arise only because $G(x,x')$ is 
%continuous across $x=x'$, see Eq. (\ref{eq:DG}) for different expressions of $\Delta G(x)$.  
In Eq. (\ref{eq:DG}) we considered $\Delta G(x)$ for specific values of $d$.  We could generalize $\Delta G(x)$ to an arbitrary $d$ 
to see if $G(x,x')$ is continuous for any other value of $d$, or the value $d=2$ is unique.  
To calculate $\Delta G(x)$ for an arbitrary $d$, we define $\Delta G$ as 
\be
\Delta G(x) =    \alpha \, \,  {\text{PV}}  \int_{-x_m}^{x_m} dy\,  \frac{p(y)}{y - x_0}.  
\label{eq:DG2}
\ee
where we used Eq. (\ref{eq:Gpm}) and the definition in Eq. (\ref{eq:DG-def}).  
"PV" in the above equations denotes the Cauchy principal value integral, accounting for the singularity at $y=x_0$.  
It ensures that infinities on the opposite sides cancel out.  

\begin{figure}[hhhh] 
 \begin{center}
 \begin{tabular}{rrrr}
 \includegraphics[height=0.19\textwidth,width=0.23\textwidth]{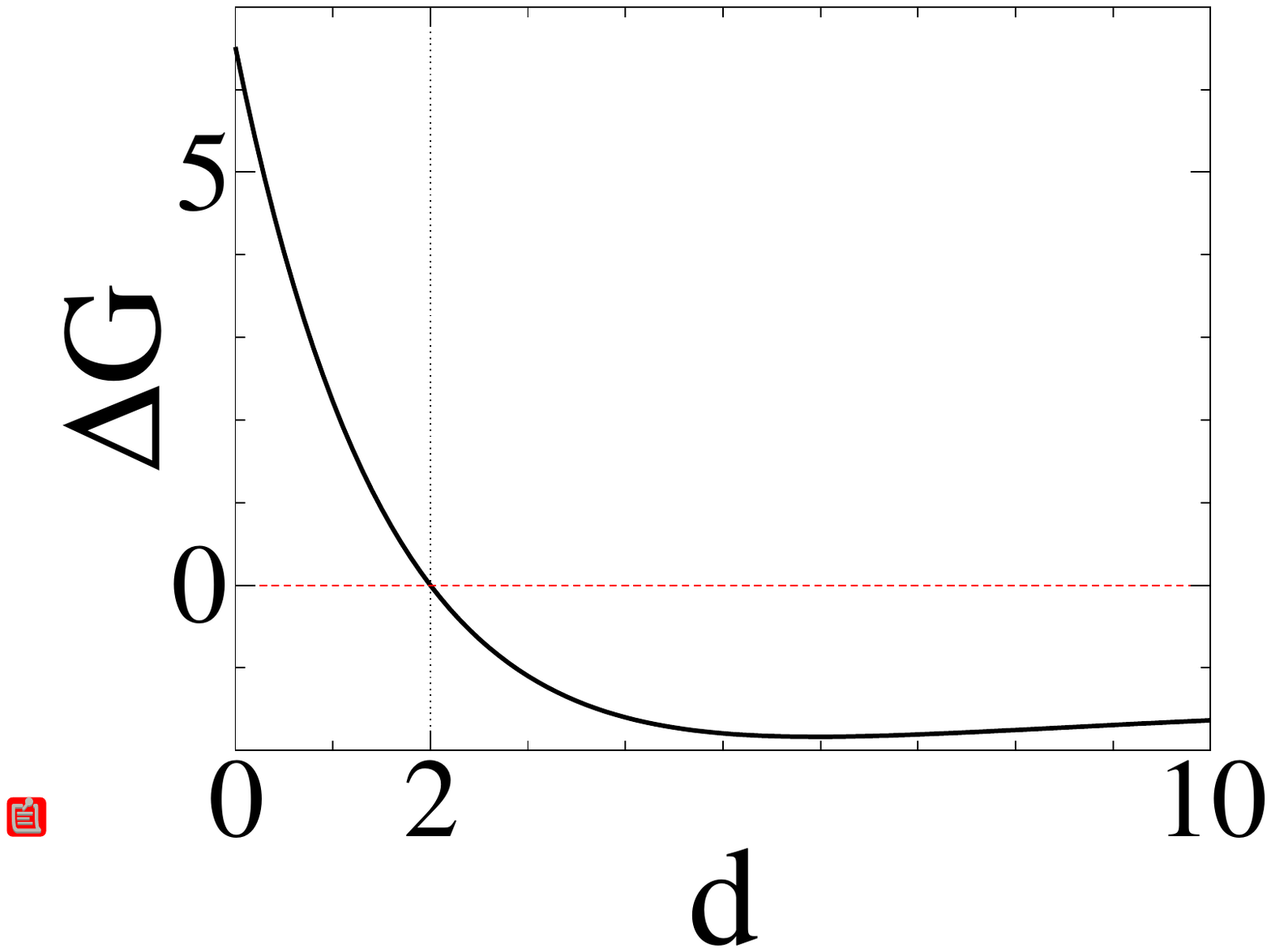} &
\end{tabular}
 \end{center} 
\caption{ The discontinuity $\Delta G(x)$, corresponding to Eq. (\ref{eq:DG3}), for $x=0.8$ and $\alpha=x_m=1$ as a function of a system dimension $d$.  } 
\label{fig:DG} 
\end{figure}

The function $p(y)$ for an arbitrary dimension $d$ can be obtained from $p_v$ considered in \cite{PRE-103-Frydel-2021}.  
This leads to the following formula 
\be
 p(y) =  \frac{1}{x_m}   \frac{\Gamma(d/2)}{ \sqrt{\pi} \, \, \Gamma(d/2 - 1/2)} \left[1 - \left( \frac{y}{x_m} \right)^2 \right]^{\frac{d-3}{2}}.  
\ee
Substituting this into Eq. (\ref{eq:DG2}) yields $\Delta G(x)$ for an arbitrary $d$:  
\be
\Delta G(x) = \frac{2 \alpha}{x_m} \frac{  \,  \Gamma(d/2) }{\Gamma(d/2-1)}    \frac{x}{x_m} \, {} _2F_1 \left( 1, 2 - \frac{d}{2}, \frac{3}{2},  \frac{x^2} {x_m^2} \right).
\label{eq:DG3}
\ee
In Fig. (\ref{fig:DG}), we plot the formula derived above as a function of $d$.  
The plot shows that $d=2$ is the only dimension for which $\Delta G(x)=0$.  
This means that generally $G(x,x')$ is irreversible.  A singular exception to this rule is for $d=2$. 
The point $d=2$ corresponds to the dimension where $\Delta G(x)$ changes sign.  
%%%%%%%%%%%%%%%%%%%%%%
\graphicspath{{figures/}}
%%%%%%%%%%%%%%%%%%%%%%%

%$$
%p_v =  \frac{1}{v_0}  \sqrt{\frac{1}{\pi}}  \frac{\Gamma(d/2)}{\Gamma(d/2 - 1/2)} \left[1 - \left( \frac{v}{v_0} \right)^2 \right]^{\frac{d-3}{2}}
%$$

\subsection{case $d=3$}
\label{sec:sec3d}

The case $d=3$ is most challenging of all dimensions.  
%Its complexity is manifest in algebraic expressions 
%of the moments $\langle x^{2n} \rangle$ of $\rho(x)$, which can be obtained by converting 
%the Fokker-Planck equation into the recurrence relation \cite{PRE-106-Frydel-2022,POF-Frydel-2023}, 
%\be
%\langle z^{2n} \rangle  =  \frac{\alpha}{2n} \sum _{k=0}^{n-1}   \frac{ \langle z^{2k}\rangle}{2n-2k+1}  \frac{(2 k+1)_{\alpha-1}}{(2 n+1)_{\alpha-1}}, 
%\label{eq:z2n-3dl}
%\ee
%%\be
%%\langle x^{2n} \rangle  =  \frac{\alpha}{2n} \sum _{k=0}^{n-1}   \frac{ \langle x^{2k}\rangle x_m^{2n-2k}}{2n-2k+1}  \frac{(2 k+1)_{\alpha-1}}{(2 n+1)_{\alpha-1}}.  
%%\label{eq:z2n-3dl}
%%\ee
%where $(a)_n = a(a-1)\dots (a-n+1)$ is the falling factorial, and $z=x/x_m$.  Because the recurrence 
%the relation cannot be solved, the moments are generated sequentially.  As in principle we can generate 
%an algebraic expression for all the moments, we should have all the information about $\rho(x)$.  
%The difficult part is how to extract $\rho(x)$ from $\langle x^{2n} \rangle$.  There is no standard 
%technique for converting moments into a distribution.  It is, therefore, interesting to see what the 
%framework based on $G(x,x')$ has to say about the case $d=3$.  
Eq. (\ref{eq:Gpm}) in this case evaluates to 
\be
G(x,x')    =    \frac{1}{2} \frac{1}{x_m}    w^{\alpha} \,\,\,  _2F_1\big( \alpha,\alpha,1+\alpha, -w\big), 
\label{eq:G3D}
\ee
where $_2F_1$ is the hypergeometric function and 
\be
w  =  \frac{ x_m } {|x - x'|}  -  \frac{x} {x - x'}.  
\ee
We have determined in Fig. (\ref{fig:DG}) that $G(x,x')$ in this dimension is discontinuous across $x=x'$ and, therefore, 
it precludes the possibility of $\rho(x)$ satisfying the detailed balance condition.  This means that we must 
extract $\rho(x)$ from Eq. (\ref{eq:integral}).

We proceed in a manner similar to that for the case $d=1$ and try to convert the integral equation in Eq. (\ref{eq:IE2}) into 
a differential equation.  This is done by differentiating Eq. (\ref{eq:IE2}) with respect to $x$.  The result is an integro-differential equation 
\ba
0    &=&   \alpha \rho \ln \left( \frac{x_m + x}{x_m - x}  \right)   +    2 x_m \rho'     \nonumber\\
&+&    \alpha \int_{-x_m}^{x} dx'\,  \frac{ \rho(x') }{ x - x'}  \left(  \frac{x_m - x}{x_m - x'}  \right)^{\alpha-1} \nonumber\\
 &+&       \alpha \int_{x}^{x_m} dx'\,  \frac{ \rho(x') }{ x - x'}  \left(  \frac{x_m + x}{x_m + x'}  \right)^{\alpha-1}.  
\label{eq:diff-3d}
 \ea
%It is not exactly a simple result, but at least it  of a hypergeometric function.   
%For the cases $d=1$ and $d=2$, 
%it was possible to convert the integral equation in Eq. (\ref{eq:integral}) into a first-order differential equation.  
%For $d=1$ because the system is very simple, and for $d=2$, due to the reversibility of the auxiliary system.  
%For the case $d=3$, obtaining a simple differential equation is no longer possible.  

To make Eq. (\ref{eq:diff-3d}) more open to interpretation, we focus on a specific case $\alpha=1$.  The
integrands in this case are simplified, and the integro-differential equation can be written as 
%\be
% j    =   -  v_0 \left[ \frac{\rho}{2} \ln \left( \frac{x_m + x}{x_m - x}  \right)      +     \frac{1}{2} \,\, PV \int_{-x_m}^{x_m } dx'\,  \frac{ \rho(x') }{ x - x'}  \right]    -   D_{eff} \rho'.  
%\ee
\be
 j    =     - \rho \frac{v_0}{2}  \ln \left( \frac{x_m + x}{x_m - x}  \right)      -     \frac{v_0}{2} \,\,  {\text{PV}}   \int_{-x_m}^{x_m } dx'\,  \frac{ \rho(x') }{ x - x'}     -   D \rho', 
\label{eq:j}
\ee
where "PV" denotes the Cauchy principal value integral accounting for the singularity at $x'=x$.  
The right-hand side of Eq. (\ref{eq:j}) is arranged to represent a flux, which for a stationary state is zero, $j=0$.  

In the original system, there is no diffusion, but in an auxiliary system, we get an effective diffusion where 
$D = v_0^2 \tau$ is the effective diffusion constant.  The first term on the right-hand side
represents an effective external force.  This force is no longer linear, as it would be in a harmonic
trap, but it traps particles into the region $x\in(-x_m,x_m)$.  The integral term 
%the term in square brackets can be interpreted as velocity due to effective forces in the system.  
%The first term can be linked to an effective external potential, and the term involving an integral 
looks a bit like a mean-field contribution of (effective) particle interactions.  The form of the integrand 
of that term suggests interactions of a log-gas \cite{Forrester}.  

The analogy with the log-gas, however, is incomplete.  To see this, below we write the flux formula of the log-gas model within the mean-field approximation, 
\be
 j_{LG}   =   - \mu \rho V'(x)      +     \frac{\mu \rho}{2} \, {\text{PV}} \int dx'\,  \frac{ \beta \rho(x') }{ x - x'}     -   D  \rho',
\ee
where $V(x)$ represents a general external potential and $\beta$ is the strength of interparticle interactions, which
for our system is $\beta = -1$.  There is one crucial difference between the mean-field term in $j_{LG}$ and that in $j$  
in Eq. (\ref{eq:j}).  The mean-field term in Eq. (\ref{eq:j}) is not proportional to $\rho(x)$.   
The absence of $\rho(x)$ in Eq. (\ref{eq:j}) introduces irreversibility into our model.
We know that the complete analogy with the log-gas model cannot be correct, because it would imply that the auxiliary system is in equilibrium, and we 
know from $\Delta G(x)$ that it is not.  

Even though the analogy is incomplete, it is still revealing, and it underscores the complexity of the problem we are 
studying.  Even if we could make a complete analogy, we could not solve the problem since there is no
general solution for the mean-field log-gas model in an arbitrary potential.  Furthermore, we can make the (incomplete)
analogy only for the case $\alpha=1$.  Any other value of $\alpha$ implies even greater mathematical complexity. 

This implies that RTP particles in simple potentials constitute their own class of mathematical models that demand
novel methods and creative approaches.

\subsubsection{numerical iteration}

Given the difficulty of exact treatment of Eq. (\ref{eq:diff-3d}), in this section we resort to 
numerical solution of Eq. (\ref{eq:integral}) via iterative procedure 
\be
\rho_{n+1}(x) = \int_{-x_m}^{x_m } dx'\, \rho_n(x') G(x,x'), 
\label{eq:rhon}
\ee
for some initial distribution $\rho_0$ that we choose to be uniform, $\rho_0 = 1/2x_m$.   
Since $\rho$ for $\alpha=0$ is uniform, we expect that the number of iterations 
necessary for $\rho_n$ to converge to $\rho$ increases with increasing $\alpha$.

\begin{figure}[hhhh] 
 \begin{center}
 \begin{tabular}{rrrr}
 \includegraphics[height=0.19\textwidth,width=0.23\textwidth]{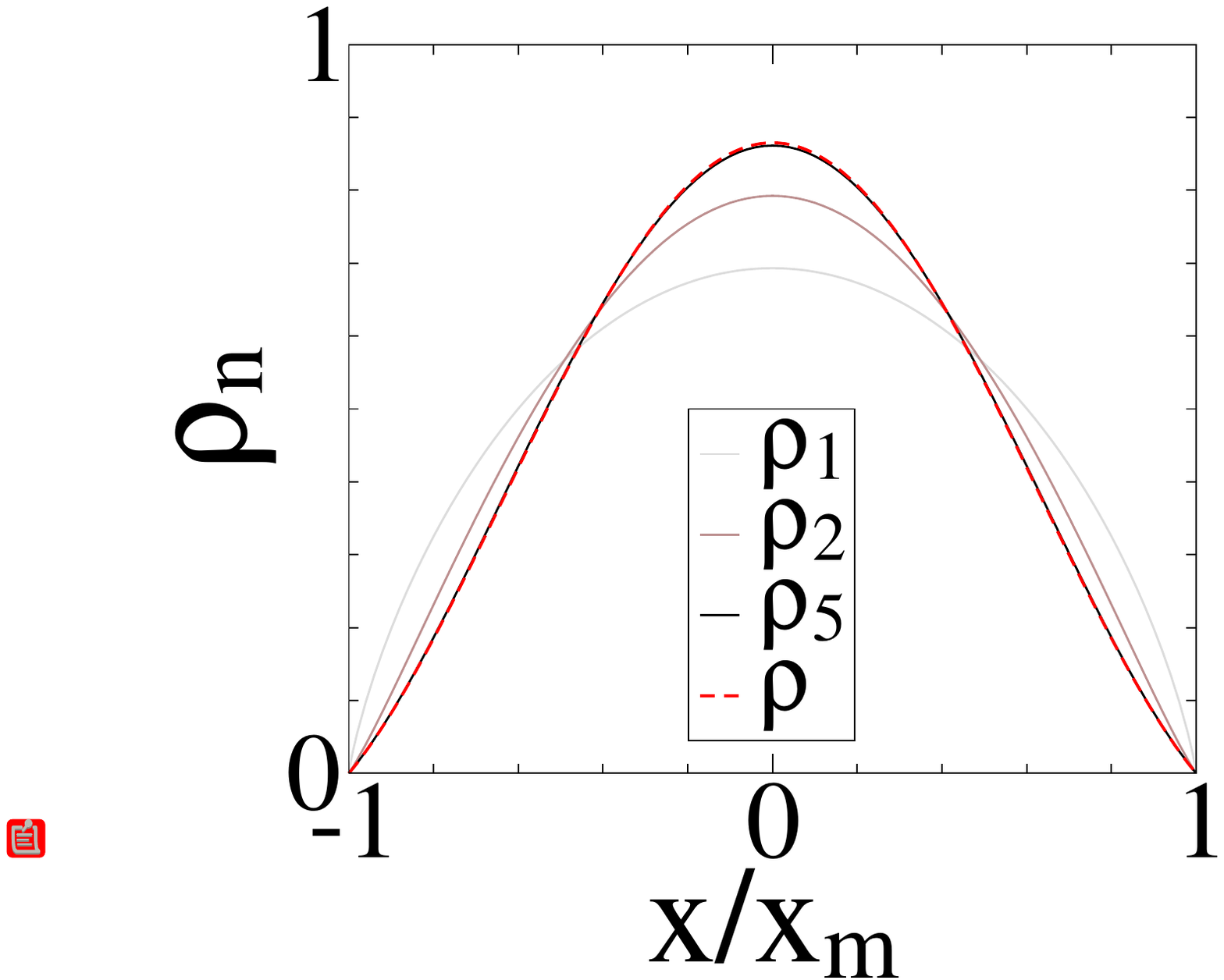} &
\end{tabular}
 \end{center} 
\caption{$\rho_{n}$ for $n=1,2,5$ obtained from numerical iterations in 
to Eq. (\ref{eq:rhon}) for the initial distribution $\rho_0=1/2x_m$ and $\alpha=1$. } 
\label{fig:rhon} 
\end{figure}

In Fig. (\ref{fig:rhon}) we plot the iterated distribution $\rho_n$ (for $\alpha=1$) for $n=1,2,5$.  
The distribution $\rho_5$ is almost indistinguishable from the exact $\rho$, indicating 
fast convergence.  
%%%%%%%%%%%%%%%%%%%%%%
\graphicspath{{figures/}}
%%%%%%%%%%%%%%%%%%%%%%%

\begin{figure}[hhhh] 
 \begin{center}
 \begin{tabular}{rrrr}
 \includegraphics[height=0.19\textwidth,width=0.23\textwidth]{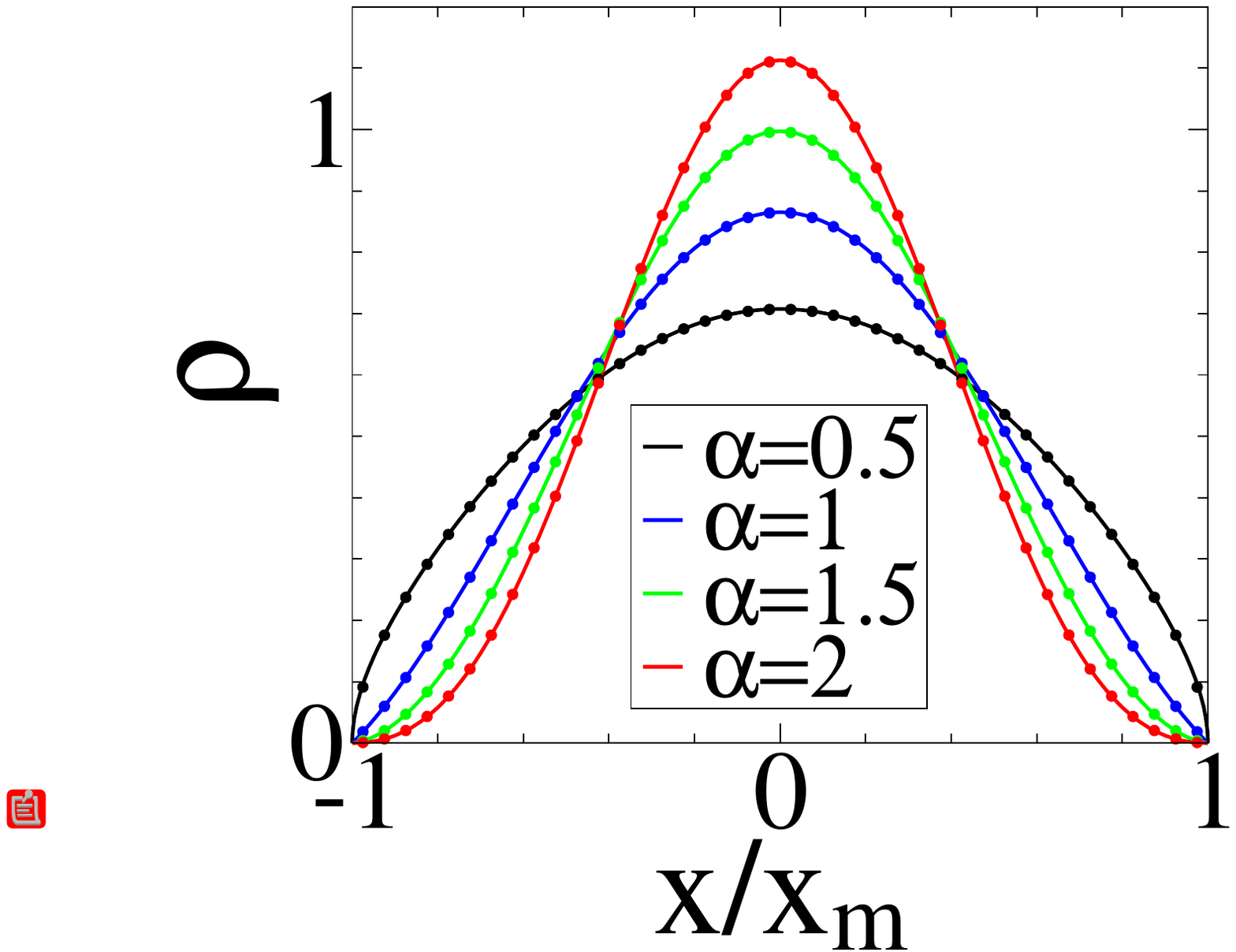} &
\end{tabular}
 \end{center} 
\caption{$\rho$ calculated from the numerical procedure in Eq. (\ref{eq:rhon}) for different 
values of $\alpha$.  The circular points represent the data points generated from a continuous 
time simulation.   
} 
\label{fig:rhon-a} 
\end{figure}

In Fig. (\ref{fig:rhon-a}) we plot a number of different distributions for different values of $\alpha$, calculated
numerically from Eq. (\ref{eq:rhon}) and compare it to $\rho$ obtained from a continuous time simulation.  
A match between the two distributions confirms the efficiency of the iterative procedure.  
%%%%%%%%%%%%%%%%%%%%%%
\graphicspath{{figures/}}
%%%%%%%%%%%%%%%%%%%%%%%

\section{Non-exponential distribution of waiting times}
\label{sec:sec2}

In the first part of this work, by mapping the stationary system of RTP particles in harmonic confinement into a jump process, 
we were able to establish a connection between the absence of discontinuity in $G(x,x')$ (which, in turn, could be connected to reversibility) 
for $d=2$ to the reduction of mathematical complexity in that particular case.  
%We were able to determine that the special status of a system in $d=2$ is the consequence of the reversibility of $G(x,x')$
%of an auxiliary system.  

But the absence of discontinuity in $G(x,x')$ alone is not a sufficient condition for reversibility.  To better understand
the emergence of reversibility, in this part of the article, we consider probability distributions of waiting times $p_t$ that are non-exponential.  
%Such reversibility could arise because $G(x,x')$ in this dimension is continuous across $x=x'$.   
The standard RTP motion assumes an exponential $p_t$.  
This is a strong condition since an exponential $p_t$ is memoryless.  It is possible that without this condition, the system in 
$d=2$ is no longer reversible and the solution no longer trivial.

%In this part of the article, we focus on a more fundamental understanding of the jump-process approach and the role of 
%the waiting times distribution $p_t$.  Specifically, we want to understand what the conditions are that make a jump-process 
%a valid sampling method, and can that method be generalized to non-exponential $p_t$.  

The hint that non-exponential $p_t$ may invalidate the jump-process approach of the previous section 
comes from the recent work in Ref. \cite{PRE-Farago-2024}.  
The authors in that work focus on RTP motion for an arbitrary $p_t$ in $d=1$.  Even for the case $d=1$, the theoretical 
framework to formulate the problem for non-exponential $p_t$ becomes non-trivial, and the resulting equations 
can only be solved numerically.  The authors of that work propose a framework involving integral equations.  It is, therefore, possible
that the framework developed for exponential $p_t$ in the previous section could be extended to non-exponential $p_t$
in the spirit of the framework in \cite{PRE-Farago-2024}.  
Apart from academic interest, as pointed out in \cite{PRE-Farago-2024,PRE-Fran-2017}, the memory effects due to a non-exponential 
$p_t$ could represent a more realistic representation of biological systems.    

%In this section, we focus on the following points: How does the jump-process algorithm depend on the distribution of waiting times? 
%If it breaks down for non-exponential distributions, can the jump-process approach be generalized to arbitrary distributions $p_t$?  

\subsection{simulation}

To investigate the role of $p_t$, we start by noting that an exponential $p_t$ corresponds to tumbling events 
occurring at a constant rate.
Thus, when considering RTP particles under confinement and in a stationary state, 
the number of tumbling events occurring at a given position is expected to be proportional to the local 
density of particles. Let $\rho_{\text{tb}}(x)$ denote the stationary, normalized distribution of tumbling events, 
and let $\rho(x)$ represent the stationary, normalized distribution of particle positions. For a process 
with a constant tumbling rate, one expects $\rho(x) = \rho_{\text{tb}}(x)$. This equality underpins the validity 
of a jump-process interpretation of RTP motion. It implies that the jump-process algorithm samples 
the distribution of tumbling events $\rho_{\text{tb}}$ (not particle positions), but due to the equality $\rho(x) = \rho_{\text{tb}}(x)$, we can use
the procedure to obtain $\rho(x)$.

For any non-exponential $p_t$—that is, any distribution that cannot be associated with a constant-rate process—
we generally expect $\rho(x) \neq \rho_{\text{tb}}(x)$. As a consequence, the straightforward implementation of the 
jump-process sampling based on the integral in Eq. (\ref{eq:integral}) breaks down. 

\begin{figure}[hhhh] 
 \begin{center}
 \begin{tabular}{rrrr}
 \includegraphics[height=0.19\textwidth,width=0.23\textwidth]{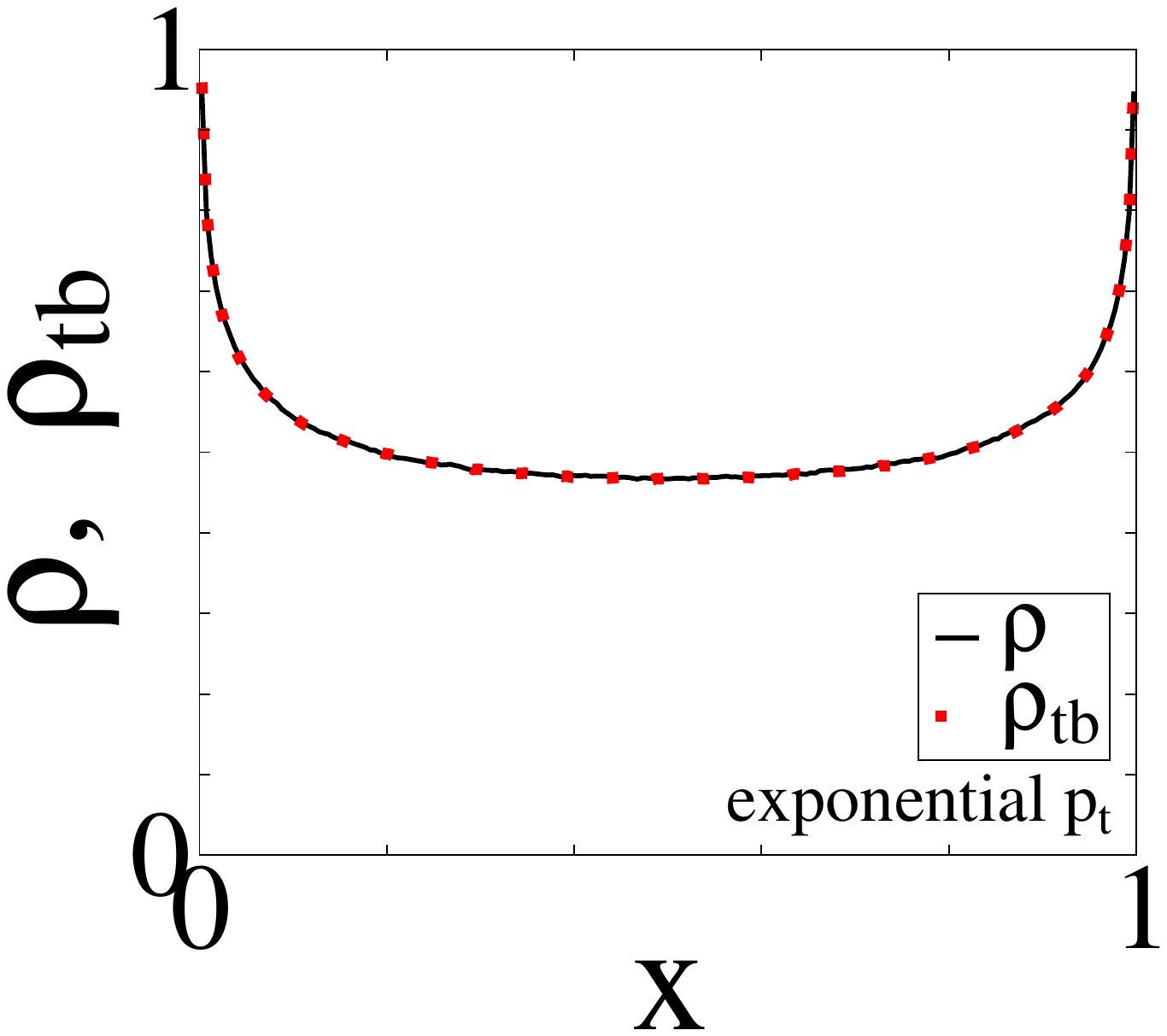} &
 \includegraphics[height=0.19\textwidth,width=0.23\textwidth]{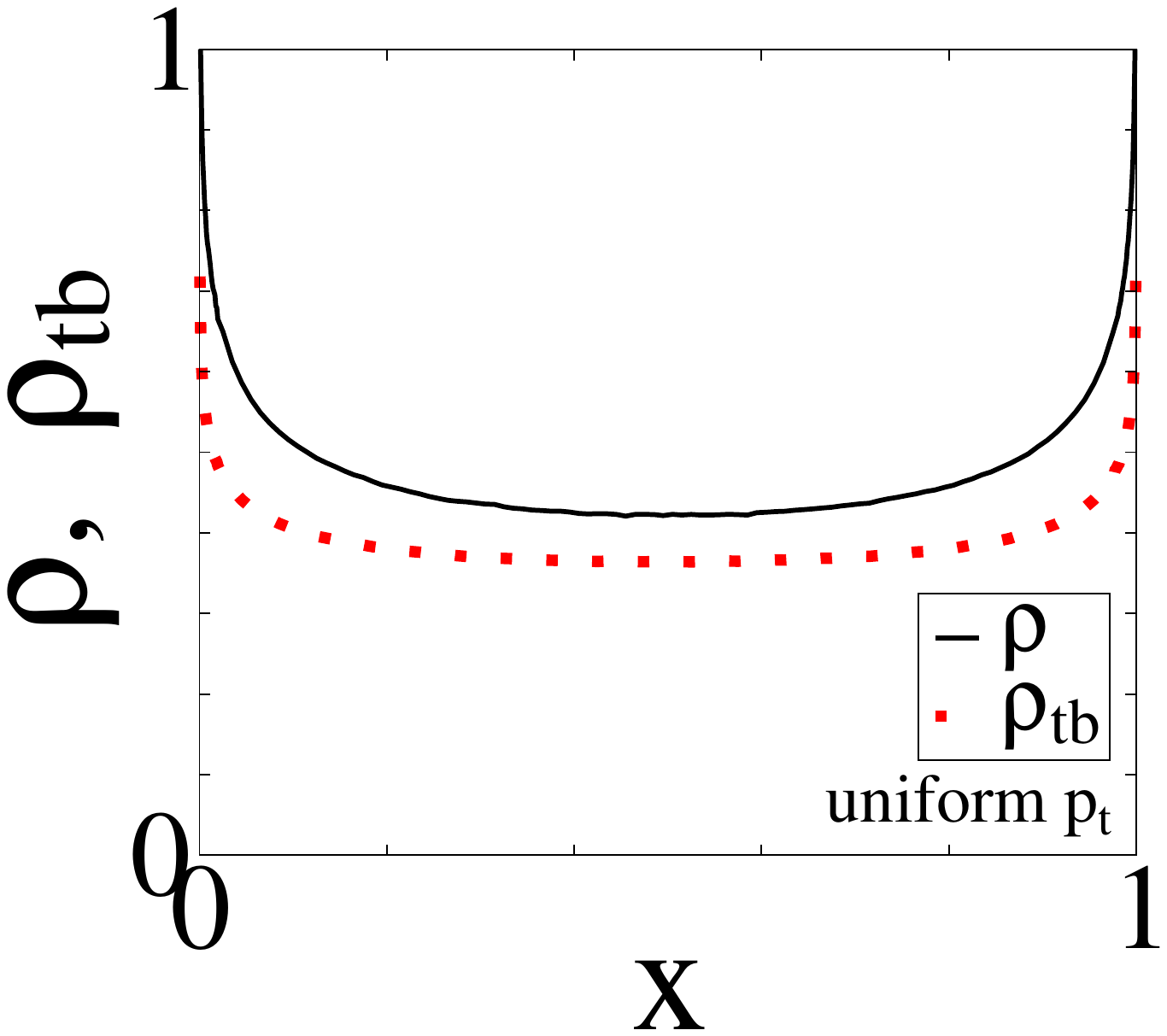} &
\end{tabular}
 \end{center} 
\caption{Stationary particle distribution $\rho(x)$ and tumbling event distribution $\rho_{\text{tb}}(x)$ for RTP 
particles confined between two parallel walls in three dimensions, obtained via simulation. 
System parameters are $L = \tau = v_0 = 1$. The left panel corresponds to exponential waiting times, 
$p_t = \tau^{-1} e^{-t/\tau}$, while the right panel corresponds to uniform waiting times, 
$p_t = (2\tau)^{-1} \Theta(2\tau - t)$. Note that because a fraction of particles becomes adsorbed 
at the walls, the distributions are not normalized to one. }
\label{fig:rho-wall} 
\end{figure}

To illustrate this point, in Fig. (\ref{fig:rho-wall}) we plot $\rho$ and $\rho_{\text{tb}}$ obtained from simulations, for
particles confined between two parallel walls in three dimensions. To obtain $\rho_{\text{tb}}$, we use the sampling 
algorithm $x_{n+1} = x_n + v t_w$ where $v$ and $t_w$ are random variables selected at each step.  To 
obtain $\rho(x)$, we use the continuous time algorithm $x(t + \Delta t) = x(t) + v \Delta t$ where $v$ remains 
constant for the duration corresponding to the waiting time $t_w$.  

In the left panel, corresponding to an 
exponential $p_t$, we observe $\rho = \rho_{\text{tb}}$, consistent with the constant-rate assumption. In contrast, 
the results for a uniform $p_t$ in the right panel clearly indicate the inequality $\rho \neq \rho_{\text{tb}}$.

\begin{figure}[hhhh] 
 \begin{center}
 \begin{tabular}{rrrr}
 \includegraphics[height=0.19\textwidth,width=0.23\textwidth]{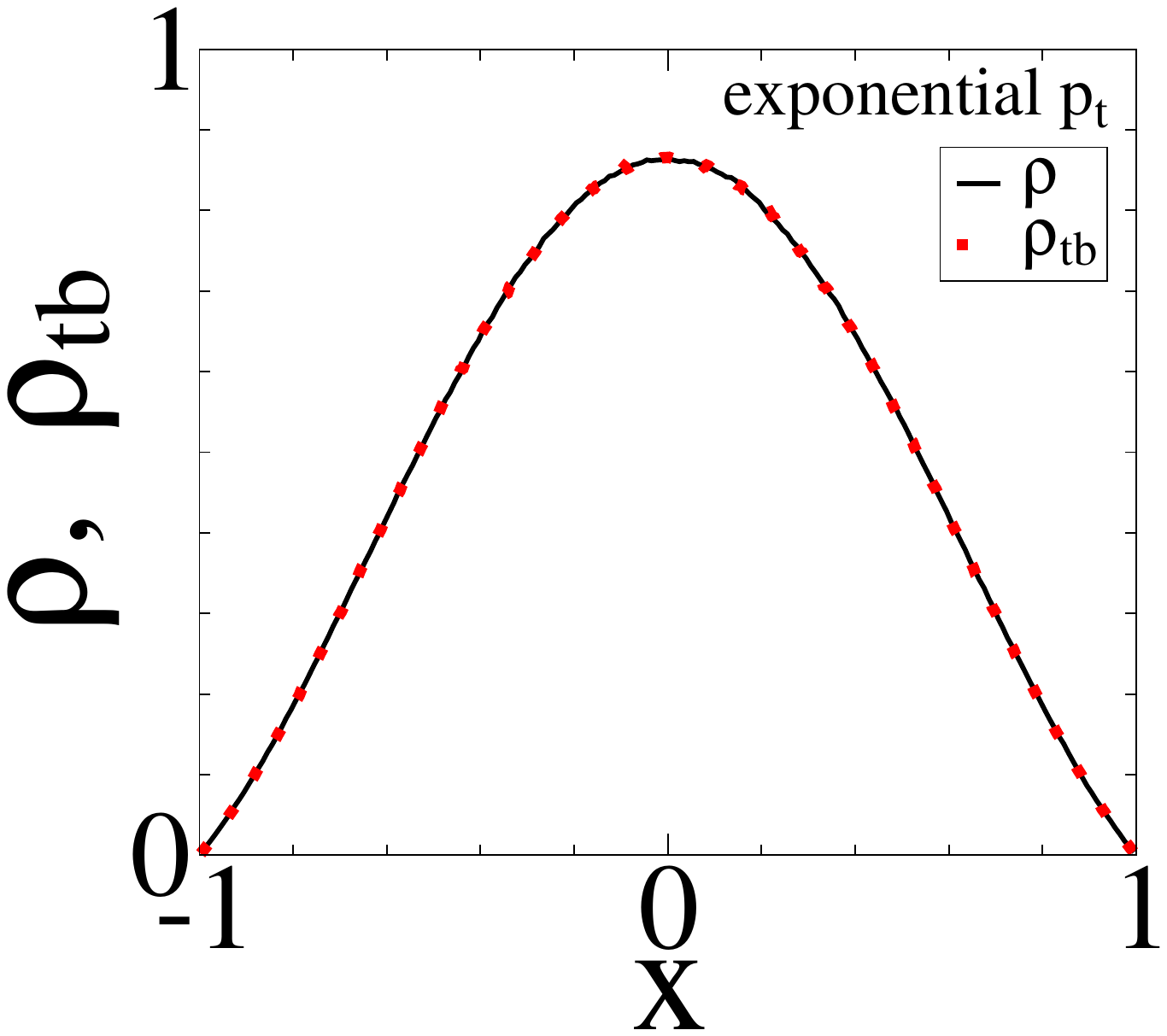} &
 \includegraphics[height=0.19\textwidth,width=0.23\textwidth]{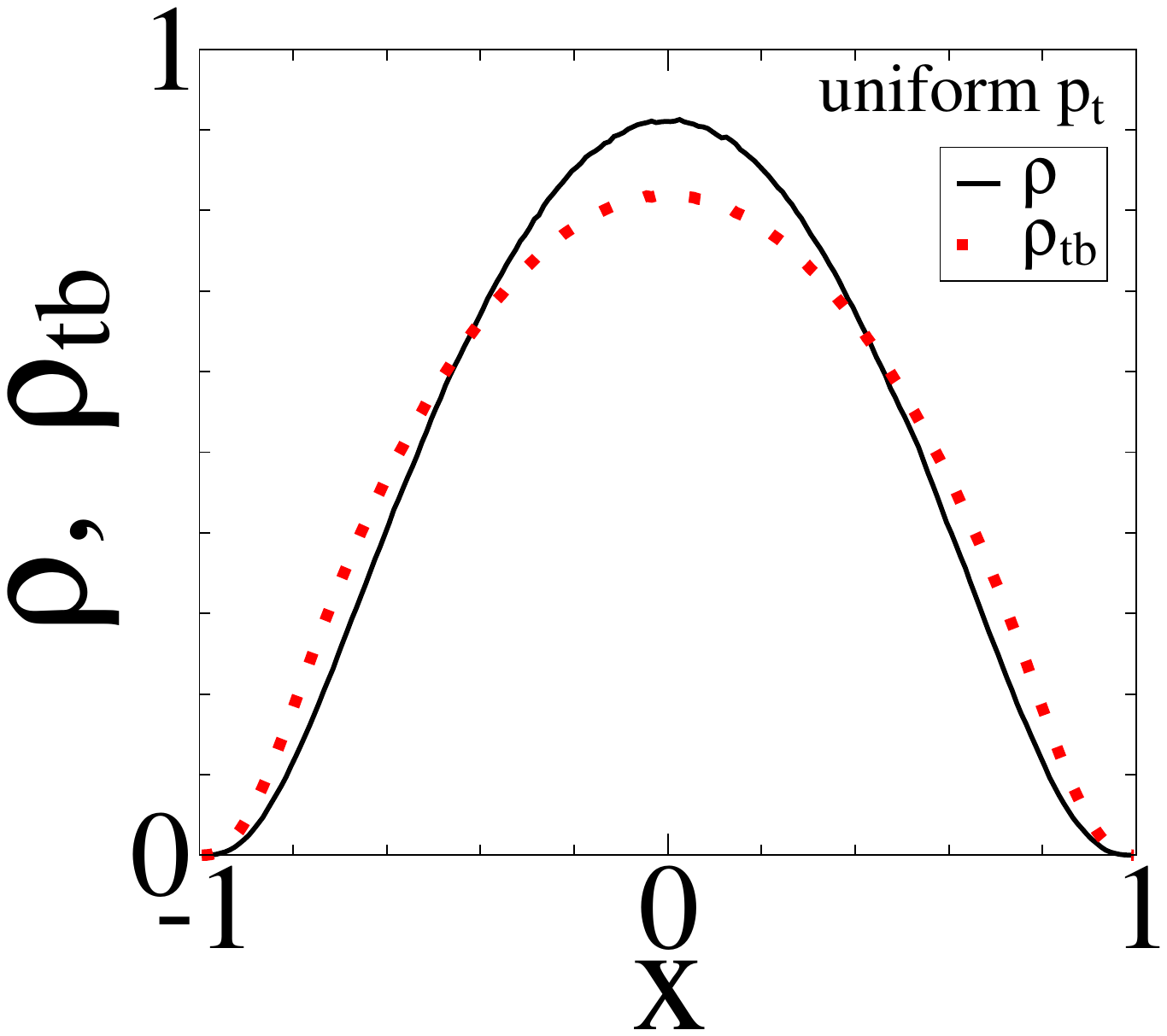} \\
\end{tabular}
 \end{center} 
\caption{An analogous plot to Fig.~\ref{fig:rho-wall}, but for RTP particles confined in a harmonic potential 
\( u(x) = \frac{Kx^2}{2} \) for the trap stiffness \( \mu K = 1 \).  As before, we compare the stationary 
particle distribution \( \rho(x) \) with the distribution of tumbling events \( \rho_{tb}(x) \), obtained from simulations.
} 
\label{fig:rho-harm} 
\end{figure}

%%%%%%%%%%%%%%%%%%%%%%%
\graphicspath{{figures/}}
%%%%%%%%%%%%%%%%%%%%%%%%
In Fig. (\ref{fig:rho-harm}), we show analogous plots for a harmonic trap. 
As in wall confinement, for an exponential \( p_t \), we find \( \rho(x) = \rho_{\text{tb}}(x) \), and 
for a uniform \( p_t \), we find \( \rho(x) \neq \rho_{\text{tb}}(x) \).
%%%%%%%%%%%%%%%%%%%%%%%
\graphicspath{{figures/}}

%To complement Fig. (\ref{fig:rho-harm}), in Fig. (\ref{fig:rhot}) we plot the ratio $\rho_{\text{tb}} / \rho$ for a uniform $p_{t}$ 
%(for an exponential $p_t$, this ratio yields one).  The plot shows a relative distribution of tumbling events, indicating 
%the greater relative density of tumbling events near the trap edges.  
%%This permits a particle to escape and find a less obstructed environment.  What is interesting 
%%is that this mechanism is an outcome of non-exponential $p_t$ without any feedback
%%mechanism.  
%%%%%%%%%%%%%%%%%%%%%%%%
%\graphicspath{{figures/}}
%\begin{figure}[hhhh] 
% \begin{center}
% \begin{tabular}{rrrr}
% \includegraphics[height=0.19\textwidth,width=0.23\textwidth]{rhot-3d-H-harm.pdf} &
%\end{tabular}
% \end{center} 
%\caption{ The ratio $\rho_{tb}/\rho$ corresponding to the right panel in Fig. (\ref{fig:rho-harm}), 
%and representing the relative density of tumbling events.} 
%\label{fig:rhot} 
%\end{figure}
%%%%%%%%%%%%%%%%%%%%%%%%%

\subsection{operator $g(x,x')$}

To understand the difference between jump-process sampling and continuous time sampling, 
%The question is, can the procedure based on discontinuous jumps be salvaged or extended to non-exponential $p_t$?  
%To help us understand the situation, 
we consider a quantity analogous to the probability distributions ot jumps $G(x,x')$ but that collects configurations not just 
at the end of the "run" stage, but continuously in time.  We designate this distribution by $g(x,x')$.  

To make things simple, we start with particles in an unconfined environment.  
Given that the motion of a particle during the "run" stage is deterministic, the particle distribution 
during this stage is represented as a propagating delta function, $p(x,x_0,t,v) = \delta(x-x_0-v t)$, 
where $x_0$ is the position at time $t=0$ and $v$ is the constant velocity.  

As we are interested in the position at the end of the "run" stage, and since the duration of the "run" stage is 
drawn from $p_t$, the delta distribution will become smeared out.  The distribution will be further smeared out 
due to the randomness of $v$, which is drawn from $p_v$ given in Eq. (\ref{eq:pv}). 
Those two contributions lead to the following distribution of jumps 
\be
G(x,x_0) = \int_0^{\infty} dt\,  p_t(t) \int_{-\infty}^{\infty} dv\, p_v(v)   \delta(x - x_0 - v t).
%p(x,t)
\label{eq:G}
\ee

Let's next look at an analogous distribution that collects configurations continuously in time.  Let's say
that we draw the time $t'$ from the distribution $p_t$.  The configurations continuously accumulated during 
the time interval $t\in [0,t']$ yield  
$
\int_0^{t} dt'\,  \int_{-\infty}^{\infty} dv\, p_v(v)   \delta(x - x_0 - v t').
$
Now, if we consider all possible $t'$ drawn from $p_t$, we get 
\be
g(x,x_0) =  \frac{1}{\tau}  \int_{0}^{\infty} dt \,   p_t(t) \int_0^{t} dt'\,  \int_{-\infty}^{\infty} dv\, p_v(v)   \delta(x - x_0 - v t').  
\label{eq:g0}
\ee
The factor $\tau^{-1}$ in front of the integral, where $\tau =  \int_{0}^{\infty} dt \,   t\, p_t(t) $, 
ensures correct normalization, $\int_{-\infty}^{\infty} dx\, g(x,x_0) = 1$.  
The formula in Eq. (\ref{eq:g0}) can be further modified using integration by parts, leading to 
\be
g(x,x_0) = \int_0^{\infty} dt\,  \frac{S(t)}{\tau}   \int_{-\infty}^{\infty} dv\, p_v(v)   \delta(x - x_0 - v t'), 
\label{eq:g}
\ee
where 
\be
S(t) = \int_t^{\infty} dt'\, p_t(t')
\ee
is the survival function.  Details of how Eq. (\ref{eq:g0}) is transformed into Eq. (\ref{eq:g}) are provided in Appendix (\ref{sec:app0b}).

Since $S(t)/\tau$ is normalized, it can be regarded as a probability distribution in time, and we 
can trace the difference between $g(x,x_0)$ in Eq. (\ref{eq:g}) and $G(x,x_0)$ in Eq. (\ref{eq:G}) 
to different distributions in time.  

Comparing Eq. (\ref{eq:g}) and Eq. (\ref{eq:G}), indicates that generally 
$G(x,x_0)\neq g(x,x_0)$, thus, the jump-process algorithm, generally, is not a valid sampling method.  
However, if 
$$
\frac{S(t)}{\tau} = p_t(t),
$$ 
then $G(x,x_0)$ and $g(x,x_0)$ are identical.  This condition is satisfied uniquely for $p_t(t) = \tau^{-1} e^{-t/\tau}$.  
For any other functional form of $p_t$ we have $S(t) \neq \tau p_t(t)$.  In those cases, the jump-process algorithm is invalid.

Similar conclusions apply to a particle in any external potential.  The equations are the same.  What is different 
is the propagating delta function; thus, to generalize Eq. (\ref{eq:g}) and Eq. (\ref{eq:G}) we use 
$$ 
\delta \left( x - x_0 - v t \right) \to \delta \left( x - x_0 - v t - \int_0^t dt'\, \mu F(x(t'))  \right), 
$$
where the term inside the delta function is obtained by integrating the Newtonian equation that 
governs particle dynamics during the deterministic "run" stage, 
\be
\dot x = v + \mu F(x), 
\label{eq:dx}
\ee
where $F(x)$ is a force due to an external potential.

\subsection{integral equation formulation for an arbitrary $p_t$}

In the section above, we defined general expressions for the distributions of jumps $G(x,x_0)$, and the distribution $g(x,x_0)$ 
that collects configurations continuously over time during each individual "run".  
In this section, we use those quantities to formulate integral equations for 
obtaining stationary distributions $\rho$ and $\rho_{\text{tb}}$, assuming non-exponential $p_t$.

First, we establish the relation between $G(x,x')$ and the distribution of tumbling events $\rho_{\text{tb}}$.  
This distribution obeys the same type of integral equation as that in Eq. (\ref{eq:integral}), 
\be
\rho_{\text{tb}}(x)  =  \int_{-\infty}^{\infty} dx'\, \rho_{\text{tb}}(x') G(x,x'),
\label{eq:rhot-A}
\ee
and as Eq. (\ref{eq:integral}), it expresses the stationarity of $\rho_{\text{tb}}$ with respect to jumps $G(x,x')$. 
To calculate the stationary distribution of particles $\rho(x)$, we use another integral equation, 
\be
\rho(x)  =  \int_{-\infty}^{\infty}  dx'\, \rho_{\text{tb}}(x') g(x,x').
\label{eq:rhot-B}
\ee
The equation above tells us that if we start with particles distributed according to $\rho_{\text{tb}}(x)$, and then 
allow those particles to perform a single "run" stage, and during that stage, accumulate configurations continuously 
in time, we will get the stationary distribution of particles $\rho$.  

It is possible to verify the validity of Eq. (\ref{eq:rhot-B}) using simulation and without explicit formulas for $G(x,x')$ and $g(x,x')$.  
%We consider the case of particles in a harmonic trap for $d=3$ and uniform $p_t$.  
Starting from some initial point in the harmonic trap, a particle makes $N$ number of jumps according to  
$x_{n+1} = x_n e^{-\mu K t_w}  +   \left( 1 -  e^{-\mu K t_w} \right) \frac{v}{\mu K}$.  The number $N$ is sufficiently 
large so that a particle's position after $N$ jumps corresponds to a random position in the distribution $\rho_{\text{tb}}(x)$.
From that position, a particle performs the last "run" stage using a time-continuous algorithm, $x(t+\Delta t) = (v - \mu K x(t)) \Delta t$.  
Particle positions are collected at each time step $\Delta t$.  These positions are used to generate $\rho(x)$.  
The two steps of the algorithm are repeated enough number of times to get good statistics.  

\begin{figure}[hhhh] 
 \begin{center}
 \begin{tabular}{rrrr}
 \includegraphics[height=0.19\textwidth,width=0.23\textwidth]{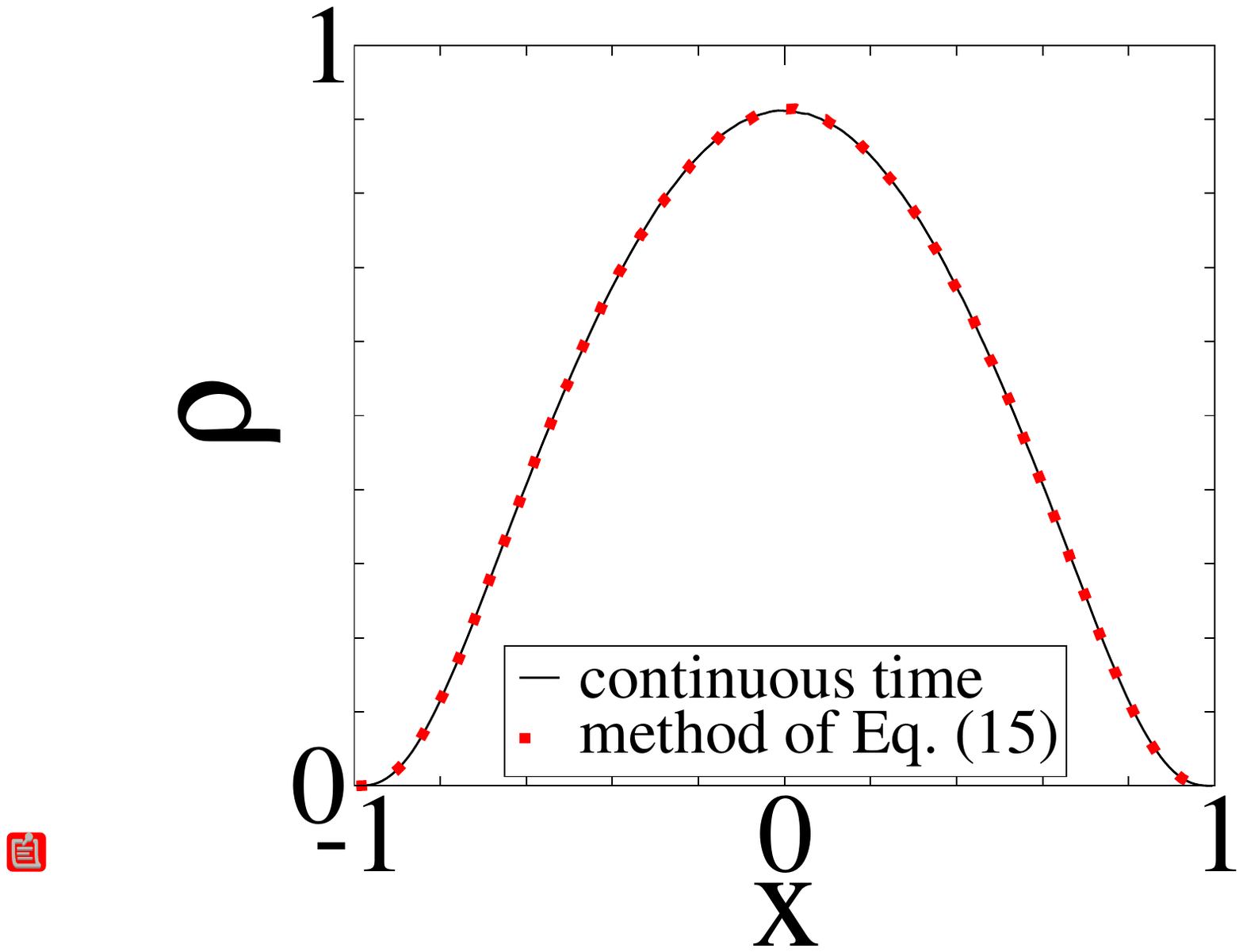} &
\end{tabular}
 \end{center} 
\caption{Verification of Eq. (\ref{eq:rhot-B}) using a simulation.  The system parameters
are the same as those in Fig. (\ref{fig:rho-harm}).} 
\label{fig:rho-B} 
\end{figure}

In Fig. (\ref{fig:rho-B}) we plot $\rho(x)$ obtained from the simulation procedure described above 
for a particle in a harmonic trap, $d=3$, and a uniform $p_t$.   
The distribution is compared with that obtained using a continuous-time simulation.  
The agreement between two distributions confirms the validity of Eq. (\ref{eq:rhot-B}).
%%%%%%%%%%%%%%%%%%%%%%%
\graphicspath{{figures/}}
%%%%%%%%%%%%%%%%%%%%%%%%

Note that for an exponential $p_t$, in which case $G(x,x')=g(x,x')$, both equations Eq. (\ref{eq:rhot-A}) and Eq. (\ref{eq:rhot-B}) 
recover the integral equation in Eq. (\ref{eq:integral}).
Eq. (\ref{eq:rhot-A}) and Eq. (\ref{eq:rhot-B}), together with the formulas for $G(x,x')$ and $g(x,x')$, 
provide a general framework for obtaining $\rho(x)$ for an arbitrary $p_t$.  

The RTP model in one-dimension and for non-exponential $p_t$ has previously been considered 
in \cite{PRE-Farago-2024}.  Our equation in Eq. (\ref{eq:rhot-A}) is analogous to Eq. (12) and Eq. (13) in that work, 
where the authors use the symbols $\Pi(x,x')$ to represents $G(x,x')$ and $i(x)$ to represent $\rho_{\text{tb}}(x)$.  
Eq. (9) in the same reference corresponds to Eq. (\ref{eq:rhot-B}) in this work, and the authors use the symbols $P(x)$ to 
represent $\rho(x)$.  The formalism in \cite{PRE-Farago-2024} is specific to a system in $d=1$, which in our framework implies 
$v = \pm v_0$.  The framework in this work is for an arbitrary distribution $p_v$.

\subsection{reversibility of $G(x,x')$ and $g(x,x')$}

While $G(x,x')$ is the transition operator associated with $\rho_{\text{tb}}(x)$,  
$g(x,x')$ is not a transition operator of $\rho(x)$.  Writing Eq. (\ref{eq:rhot-B}) in a self-consistent form, 
\be
\rho(x)  =  \int_{-\infty}^{\infty}  dx'\, \rho(x') \left[ \frac{\rho_{\text{tb}}(x')}{\rho(x')} g(x,x') \right],
\ee
allows us to identify $g_{eff}(x,x') = {\rho_{\text{tb}}(x')}{} g(x,x') / \rho(x')$ as the transition operator associated with $\rho(x)$.  

It can be easily shown that $g_{eff}(x,x')$ satisfies Kolmogorov loop criterion in Eq. (\ref{eq:KC1}) (indicating that $g_{eff}(x,x')$ is reversible) 
only if $g(x,x')$ does.  But as the detailed balance condition resulting from such a reversibility, $\rho(x') g_{eff}(x,x') = \rho(x) g_{eff}(x',x)$, 
can be simplified to $\rho_{\text{tb}}(x') g(x,x') = \rho_{\text{tb}}(x) g(x',x)$, it would mean that we cannot have at the same time 
$\rho_{\text{tb}}(x') G(x,x') = \rho_{\text{tb}}(x) G(x',x)$ since for non-exponential $p_t$, $g(x,x') \neq G(x,x')$.
This means that the possibility of both $G(x,x')$ and $g(x,x')$ being reversible is inconsistent within the framework.

We next show that $g(x,x')$ cannot be reversible, as this would lead to another inconsistency. 
The reversibillity of $g(x,x')$ implies that we could represent $\rho_{\text{tb}}(x')$ as
$
\rho_{\text{tb}}(x') = \rho_{\text{tb}}(x) {g(x',x)} / {g(x,x')}.  
$
But if we substitute this into Eq. (\ref{eq:rhot-B}) we get  
$$
\rho(x)  =  \rho_{\text{tb}}(x) \int_{-\infty}^{\infty}  dx'\,  g(x',x),
$$
and since $g(x,x')$ is normalized, $\int_{-\infty}^{\infty}  dx'\,  g(x',x) = 1$, we get
$
\rho(x)  =  \rho_{\text{tb}}(x),
$
which for non-exponential $p_t$ is inconsistent.  Thus, we can exclude the possibility of $g(x,x')$
being reversible.  This leaves $G(x,x')$ as the only possible operator that could be reversible.

\subsection{$\Delta G(x)$ for an arbitrary $p_t$}

We next focus on the operator $G(x,x')$ and the possibility of it being reversible.  It was previously determined that $G(x,x')$ 
cannot be reversible if it is discontinuous across $x=x'$.  We have quantified the discontinuity at a given $x$ as $\Delta G(x)$
defined in Eq. (\ref{eq:DG-def}).   Only for $d=2$ the discontinuity was found to be zero.  Here, we want to 
understand how the non-exponential $p_t$ affects the behavior of $\Delta G(x)$.

It can be shown that the function of discontinuity for an arbitrary $p_t$ is given by 
\be
\Delta G(x) =    \frac{p_t(0)}{\mu K} \, \,  {\text{PV}}  \int_{-x_m}^{x_m} dy\,  \frac{p(y)}{y - x_0}.  
\label{eq:DG2A}
\ee
This result is a more general version of the formula Eq. (\ref{eq:DG2}), which was derived specifically for an exponential $p_t$.  
Since for an exponential distribution, $p_t(0) = \tau^{-1}$, Eq. (\ref{eq:DG2A}) recovers Eq. (\ref{eq:DG2}).  
The dependence of $\Delta G(x)$ on $p_t(0)$ is intuitively correct, since discontinuity arises in the neighborhood of the initial 
point, $x=x'$, thus, it depends on the probability that the waiting time is zero.  

Using Eq. (\ref{eq:DG2A}), we can arrive at a generalized result of Eq. (\ref{eq:DG}), 
\be
 \Delta G(x)
=  
\begin{cases}
      \frac{p_t(0)}{\mu K} \frac{x}{x_m^2 - x^2} , & \text{for $d=1$},\\
      0, & \text{for $d=2$}, \\
       \frac{p_t(0)}{\mu K}  \frac{1}{2 x_m} \ln \big( \frac{x_m + x }{x_m - x} \big) , & \text{for $d=3$}. \\
  \end{cases}
\label{eq:DGH}
\ee
From Eq. (\ref{eq:DGH}) we can tell that a non-exponential $p_t$ cannot create a discontinuity in $G(x,x')$ for $d=2$, 
suggesting that the absence of discontinuity in this dimension is a particular property associated with that dimension.

\subsection{Uniform $p_t$}

The absence of discontinuity does not guarantee that $G(x,x')$ is reversible.  It is a necessary but insufficient condition of reversibility.  
In this section, we consider a uniform $p_t$, where it can be demonstrated in a straightforward manner 
that $G(x,x')$ is irreversible.  

For a uniform distribution given by $p_t = (2\tau)^{-1} \Theta(2\tau - t)$, 
one can use Eq. (\ref{eq:xtp}) to demonstrate that the maximum distance from a trap center that a particle can reach (assuming the initial position at $x_0=0$) is when $v=v_0$ and $t=2\tau$, 
and is given by 
$$
x_{\text{max}}    =       \left( 1  -  e^{- 2\mu K \tau} \right) x_m,
$$
%\be
%x_{max}    =       \left( 1  -  e^{- 2/\alpha} \right)  x_m,
%\ee
so that $x_{\text{max}} < x_m$.
This means that the probability distribution of jumps for the initial position $x_0=0$ vanishes for $x>x_{\text{max}}$, 
$$
G(x > x_{\text{max}} ,0) = 0.  
$$

On the other hand, for a particle initially at $x=x_{\text{max}}$, the minimum distance is 
%\be
%x(t)    =    x_{\text{max}} e^{- 2\mu K \tau}    -    \left( 1  -  e^{-2\mu K \tau } \right) x_m,
%\ee
%\be
%x(t)    =    \left( e^{- 2\mu K \tau}  -  e^{- 4\mu K \tau} \right) x_m    -   \left( 1  -  e^{-2\mu K \tau } \right) x_m,
%\ee
%\be
%x(t)    =    -\left( -2e^{- 2\mu K \tau}  + e^{- 4\mu K \tau}  +  1  \right) x_m   
%\ee
$$
x_{\text{min}}    =    -\left(   1  -  e^{- 2\mu K \tau}   \right)^2 x_m.
$$
This means that while a particle initially at $x=0$ cannot reach any point for $x>x_{\text{max}}$.  
There are positions $x>x_{\text{max}}$ that can reach the position $x=0$, 
$$
G(0,x>x_{\text{max}})>0.
$$  
Given that $\rho_{\text{tb}}(x)$ is defined on the support $x\in(-x_m,x_m)$, this means that 
the detailed balance relation between the points $x=0$ and $x_m >y>x_{\text{max}}$, 
$$
\rho_{\text{tb}}(y) G(0,y)  = \rho_{\text{tb}}(0) G(y,0),
$$
could not be satisfied.  The assumption of reversibility leads to inconsistencies.  
Consequently, we conclude that $G(x,x')$ for a uniform $p_t$ is irreversible even for $d=2$ where $\Delta G = 0$, 
confirming that $\Delta G = 0$ is a necessary but insufficient condition of reversibility.

%By assuming that $G(x,x')$ is reversible, we should be able to represent $\rho_{\text{tb}}(x)$ as 
%\be
%\rho_{\text{tb}}(x) \propto \frac{G(x,0)}{G(0,x)}, 
%\label{eq:rhotb-H}
%\ee
%see Eq. (\ref{eq:rho-DB}) for a similar expression.  According to Eq. (\ref{eq:rhotb-H}), $\rho_{\text{tb}}(x>x_{\text{max}}) = 0$.  
%However, we know that $\rho_{\text{tb}}(x)$ is defined on the inteval $x\in(-x_m,x_m)$.  For example, for a particle that 
%starts at $x_0=x_{\text{max}}$, the maximal position when $t=2\tau$ and $v=v_0$ is twice the maximal position in Eq. (\ref{eq:xmax}), 
%$x   =   2 x_{\text{max}}$.  This means that there should be sequences of jumps that can take a particle to $x=x_m$.  

\subsection{superposition of exponential distributions}

A simple way to construct a memoryless distribution is to use a mixture of exponential distributions, 
\be
p_t = \sum_{n=1}^N \frac{p_n}{\tau_n} e^{-t/ \tau_n}, 
\label{eq:pt-mix}
\ee
where $\sum_{n=1}^N p_n = 1$.  The transition operator of such $p_t$ is a sum of operators in Eq. (\ref{eq:G2D})
for different $\alpha$, 
\be
G(x,x')    \equiv    \sum_{n=1}^N p_n G(x,x'; \alpha_n),
%p G(x,x'; \alpha_1)    +   (1-p) G(x,x'; \alpha_2).  
\label{eq:G2D-mix}
\ee
where $\alpha_n = (\mu K \tau_n)^{-1}$.  
By considering the simplest form of a memoryless $p_t$, a mixture of two exponential distributions 
\be
p_t = p \tau_1^{-1} e^{-t/\tau_1} + (1-p) \tau_2^{-1} e^{-t/\tau_2},
\label{eq:pt-2}
\ee  
where $p\in (0,1)$ and $\tau_1\neq \tau_2$,  it can be easily shown, using Eq. (\ref{eq:G2D}) and Eq. (\ref{eq:G2D-mix}), 
that the resulting $G(x,x')$ is irreversible.

The same conclusion applies to any other mixture of exponential distributions defined in Eq. (\ref{eq:pt-mix}).  
The same conclusion applies to continuous superpositions of exponential distributions
$$
p_t = \int d\tau \, \frac{p(\tau)}{\tau} e^{-t/ \tau}.  
$$
%$$
%G(x,x')    \equiv    \int d\alpha \, p(\alpha) G(x,x'; \alpha),
%$$

This means that for a class of distributions that can be represented as a superposition of exponential distributions, 
the resulting operator $G(x,x')$ is irreversible.  Distributions included in this class are those that are completely 
monotonic.   This includes Gamma distributions and inverse Gaussian distributions, but does not include a 
semi-Gaussian or a half-t distribution.

%and hypoexponential distribution 
%$$
%p_t    =    \frac{ e^{-t/\tau_1}  -  e^{-t/\tau_2}  }{\tau_1 - \tau_2}  
%$$
%For the case $d=2$, the formula in Eq. (\ref{eq:Gpm}) evaluates to 
%\ba
%G(x,x') &=& \frac{ \Gamma(\alpha+1) } { 2^{\alpha} \sqrt{\pi} \Gamma(\alpha+1/2)} \sqrt{\frac{1}{x_m^2 - x'^2}} \left( \frac{x_m^2 - x^2}{x_m^2-x'^2} \right)^{\frac{\alpha}{2} - \frac{1}{2}}
%\nonumber\\ 
%&&
%w^{\alpha/2}   \,\,\,     _2F_1 \left( \frac{\alpha}{2}, \frac{\alpha}{2} + \frac{1}{2}, \alpha + \frac{1}{2}, w   \right),
%\ea
%where $_2F_1$ is the hypergeometric function.  To simplify the nomenclature, we introduce 
%a dimensionless parameter 
%$$
%w = \left[1 + \frac{ (x-x')^2 x_m^2} { (x_m^2-x^2) (x_m^2 - x'^2)} \right]^{-1}.  
%$$

\subsection{arbitrary $p_t$}

For any other form of $p_t$, to show that $G(x,x')$ is irreversible in 2D, we write Eq. (\ref{eq:G}) as
\be
G(x,x') = \int_0^{\infty} dt\,  p_t(t)  p(x,x',t), 
\label{eq:G2DA}
\ee
where the analytical expression of the distribution $p(x,x',t)$ for $d=2$ is 
%$$
%p(x,x_0,t) =  \int_{-\infty}^{\infty} dv\, p_v(v)   \delta(x - x_0 - v t)
%$$
%$$
%p(x,x_0,t) =  \int_{-\infty}^{\infty} dv\, p_v(v)   \delta\left(x - x_0 e^{-\mu K t_w}  -   \left( 1 -  e^{-\mu K t_w} \right) \frac{v}{\mu K} \right)
%$$
%$$
%p(x,x_0,t)   =   \frac{\mu K }{ | 1 -  e^{-\mu K t} | }
%\int_{-\infty}^{\infty} dv\, p_v(v)    \delta\left(  v  -   \mu K \frac{ x - x_0 e^{-\mu K t} } { 1 -  e^{-\mu K t} }  \right)
%$$
%\be
%p_v    =   \frac{1}{\pi}   \frac{1}{\sqrt{v_0^2-v^2}}, 
%\ee
%$$
%p(x,x_0,t)   =   \frac{1 }{ | 1 -  e^{-\mu K t} | }  \frac{1}{\pi}   \frac{1}{\sqrt{x_m^2 -  \left(\frac{ x - x_0 e^{-\mu K t} } { 1 -  e^{-\mu K t} }\right)^2 }  }  
%$$
\be
p(x,x',t)   =   \frac{1}{\pi}   \frac{1}{\sqrt{x_m^2(1 -  e^{-\mu K t} )^2 -  \left(  x - x' e^{-\mu K t}  \right)^2 }  }, 
\label{eq:pxyt}
\ee
with the support
$$
x\in \bigg[ y e^{-\mu K t}   -  x_m(1 - e^{-\mu K t} ),  y e^{-\mu K t}   +   x_m(1 - e^{-\mu K t} ) \bigg].  
$$
By integrating Eq. (\ref{eq:G2DA}) numerically for an arbitrary $p_t$, we can then test the resulting $G(x,x')$ for reversibility.  
We found that only an exponential $p_t$ leads to a reversible $G(x,x')$.

What is interesting about $p(x,x',t)$ in Eq. (\ref{eq:pxyt}) is that at no point in time $t$ this distribution is 
reversible.  It can only become reversible after it is integrated over time, as in Eq. (\ref{eq:G2DA}), and only 
if $p_t$ in the integrand is exponential.  We conclude that for the operator $G(x,x')$ to be reversible, 
it must not have any discontinuities, and the distribution of waiting times $p_t$ must be memoryless.  
The two conditions are only met for a system in 2D with exponential $p_t$.

\section{Conclusion}
\label{sec:sec4}

By reinterpreting RTP motion as a jump process, we arrive at an integral equation formulation 
of a system in a stationary state.    
The formulation is achieved by mapping RTP motion into a jump process.  
The key quantity of the formulation is the transition operator $G(x,x')$ and the integral equation $\rho(x) = \int dx'\, \rho(x') G(x,x')$.  
The resulting framework is used to analyze the system of RTP particles in a harmonic potential in different dimensions.  
The special focus is on a system in 2D, for which an exact closed-form solution is obtained.  

No closed-form solution is possible for dimensions $d>2$.  The special status of a system in 2D 
is attributed to the fact that the transition operator $G(x,x')$ of an auxiliary system is reversible.  
The reversibility in this case is linked to the absence of the discontinuity in $G(x,x')$ at $x=x'$.  
The absence of discontinuity, however, is a necessary but not sufficient condition of reversibility.  It is determined 
that for a non-exponential distribution of waiting times $p_t$, the corresponding $G(x,x')$ is irreversible.

By extending the integral equation framework to non-exponential $p_t$, Eq. (\ref{eq:rhot-A}) and Eq. (\ref{eq:rhot-B}), 
it can be determined from that framework alone, without considering a specific form of $G(x,x')$ and $g(x,x')$, 
that the stationary distribution $\rho(x)$ of the auxiliary system can never be interpreted as an equilibrium 
distribution.  

As regards the system in 3D, even for an exponential $p_t$, it constitutes an inherently complex problem.  
For the case $\alpha=1$ we were able to make an incomplete analogy between such a system and the 
log-gas model in the mean-field approximation.

Since the mapping of RTP particles in confinement to the jump-process algorithm depends on omitting
the deterministic "run" stage governed by the Newtonian equation
$
\dot x = v - \mu F(x),
$
and the subsequent calculation of the kernels $G(x,x')$ and $g(x,x')$, it is suitable for the cases which admit 
of analytical solutions to the Newtonian equation.  
In \cite{POF-Frydel-2024}, it was applied to $F = 0$ and in the current work to $F = -K x$.  The method could be 
more challenging for more complex $F(x)$.  It possibly could be applied to the case of a periodic force $F \propto \cos(x)$
up till now not explored.  For more complex $F(x)$ it might require a combination of numerical and analytical procedures.

This work highlights the mathematical complexity of self-propelled particles, even for simple 
potentials and in a stationary state.  These systems constitute their own class of mathematical problems.  
As such, they are worthy of further investigation. There still remains significant work to better understand these 
systems.  There is plenty of room for creativity and application of novel techniques.   
A better understanding of those systems might contribute to a broader understanding of statistical mechanics in general.

\begin{acknowledgments}
D.F. acknowledges financial support from FONDECYT through grant number 1241694.  
\end{acknowledgments}

\section{DATA AVAILABILITY}
The data that support the findings of this study are available from the corresponding author upon 
reasonable request.

\vspace{1cm}
\appendix

\section{Fokker-Planck equation for a general dimension}
\label{sec:app0}

In this section we generalize the Fokker-Planck equation for RTP particles in 
confinement given in Eq. (\ref{eq:FP2D}) specifically for dimension $d=2$.  
A general form of this equation, for a dimension $d>1$, could be written as 
%The idea is that for dimensions $d>1$, the integral in the second term of Eq. (\ref{eq:}) is over 
%a solid angle.  
\ba
\frac{\partial n}{\partial t} &=& -\bnabla\cdot \left[ \left({\bf F}  +  v_0 {\bf u} \right) n \right] 
-    \frac{ 1 }{\tau} \left(  n  -   \frac{1}{ \int d\Omega } \int d\Omega \, n \right)
\label{eq:app0-2}
\ea
%\be
%\frac{\partial n}{\partial t} = -\bnabla\cdot \left[ \left({\bf F}  +  v_0 {\bf u} \right) n \right]  
% -    \frac{ 1 }{\tau} \left(  n  -   \frac{1}{2\pi} \int_{0}^{2\pi} d\theta\, n({\bf r},\theta,t)\right),
%\label{eq:app0-1}
%\ee
where $\Omega$ is a solid angle.  The integral considers all orientations of 
a swimming velocity whose magnitude is constant $|{\bf v}_{swim}| = v_0$.  
For dimension $d=3$, where $d\Omega = d\theta d\phi \sin\phi$, the Fokker-Planck equation
becomes
\ba
\frac{\partial n}{\partial t} &=& -\bnabla\cdot \left[ \left({\bf F}  +  v_0 {\bf u} \right) n \right]  \nonumber\\ 
&-&    \frac{ 1 }{\tau} \left(  n  -   \frac{1}{4\pi} \int_{0}^{2\pi} d\theta \int_0^{\pi} d\phi \, \sin\phi n({\bf r},\theta,\phi,t)\right)
\label{eq:app0-3}
\ea
where $n\equiv n({\bf r},\theta,\phi,t)$.  

For the system in $d=1$, there are only two possible swimming orientations.  In this case, the distribution of particles 
with forward and backward direction of motion are represented as $n_+$ and $n_-$, respectively, 
and the Fokker-Planck equation is represented as two coupled differential equations:
\ba
&& \frac{\partial n_+}{\partial t} = -\bnabla\cdot \left[ \left({F}  +  v_0 \right) n_+ \right]    -    \frac{ 1 }{\tau} \left(  n_+  -   n_- \right)  \nonumber\\
&& \frac{\partial n_-}{\partial t} = -\bnabla\cdot \left[ \left({F}   -   v_0 \right) n_+ \right]    -    \frac{ 1 }{\tau} \left(  n_-  -   n_+ \right).  
\label{eq:app0-4}
\ea

\section{Derivation of the probability distributions $p_v$}
\label{sec:app0a}

To obtain the probability distribution for the projection of the swimming velocity along the 
$x$-axis, $v = v_0 \cos\theta$, we start by considering the fact that in the polar coordinates, 
the swimming velocity is uniformly distributed over the angle $\theta$.  Such a uniform 
distribution on the interval $\theta \in [0,2\pi]$ is given by $p_{\theta} = 1/(2\pi)$.  The projection of the 
swimming velocity along the $x$-axis will be distributed on the interval $v\in[-v_0,v_0]$.  To obtain
this distribution, we start with the integral $\frac{1}{2\pi} \int_0^{2\pi} d\theta \, p_{\theta}(\theta)$
and then apply a change of a variable, 
\be
v = v_0 \cos\theta, ~~~ dv = -v_0 \sin\theta d\theta  = - d\theta \sqrt{ v_0^2 - v^2}.  
\label{eq:app0a-1}
\ee
This yields 
\be
\frac{1}{2\pi} \int_0^{2\pi} d\theta  =  -\frac{1}{\pi} \int_{-v_0}^{v_0} d v\, \frac{d\theta}{d v} 
= \frac{1}{\pi} \int_{-v_0}^{v_0} dv\, \frac{1}{\sqrt{v_0^2-v^2}}.
\ee
From the final expression, we conclude that the distribution in $v$ is
\be
p_v = \frac{1}{\pi}  \frac{1}{\sqrt{v_0^2-v^2}}.
\ee
This expression appears in Eq. (\ref{eq:pv}) and corresponds to the case $d=2$. 

In the case $d=3$, we work with spherical coordinates.  In this case, the orientation 
of any vector is parametrized by two parameters, $\theta$ and $\phi$.  The orientation
of a swimming orientation is uniformly distributed over a solid angle $\Omega(\theta,\phi)$.  
And because in spherical coordinates the integral over a solid angle is 
$ \int_0^{2\pi} d\phi \int_0^{\pi} d\theta\, \sin\theta$, we start from this integral 
and then transform it into the integral over $v$ using a change of variable in Eq. (\ref{eq:app0a-1}).  
This yields 
\be
\frac{1}{4\pi} \int_0^{2\pi} d\phi \int_0^{\pi} d\theta\, \sin\theta   
=
- \frac{1}{2} \int_{-v_0}^{v_0} d v \,  \frac{d\theta}{d v} \sin\theta   
= 
\frac{1}{2}\frac{1}{v_0} \int_{-v_0}^{v_0} dv.  
\ee
From the final expression, we conclude that the distribution in $v$ is uniform on the 
interval $v \in [-v_0, v_0]$:  
\be
p_v = \frac{1}{2} \frac{1}{v_0}.  
\ee
This expression appears in Eq. (\ref{eq:pv}).

%\appendix
%\section{Jump-algorithm simulation}
%\label{sec:app-sim}
%
%\begin{lstlisting}%[basicstyle=\small]
%
%import math
%import random
%import numpy as np
%import time
%from scipy import stats
%
%tau=1
%v0=1
%K=1
%
%NT  = 100000
%
%x = 0
%for i in range(0,NT):
%	v = v0*np.random.uniform(-1,1)
%        tp = np.random.uniform(0,2*tau) 
%
%        ts = tp*np.random.uniform(0,1)
%        xs = (x-vx/K)*math.exp(-K*ts) + vx/K
%
%        x  = (x-vx/K)*math.exp(-K*tp) + vx/K
%        
%        count += 1 
%        if i%1000000==0 and i>1:
%                print(i,x,vx,tp,ts,sum2/i,sum2s/i,sum4s/i)  
%        k = int((xs+xm)/bin)
%        if k<NBIN and k>=0:
%                p[k] += tp/tau
%                
%\end{lstlisting}

\section{ Derivation of Eq. (\ref{eq:g})}
\label{sec:app0b}

In the main body of this work, the expression in Eq. (\ref{eq:g0}) is converted to Eq. (\ref{eq:g}).  
In this section we provide more details of this transformation.  
For clarity we repeat Eq. (\ref{eq:g0}) below
\be
g(x) =   \frac{1}{\tau}  \int_{0}^{\infty} dt_w \,   p_t(t_w)  \left[\int_0^{t_w} dt\, p(x,t)\right].  
\ee
Using integration by parts, we express the above integral as 
\ba
g(x)  &=&   -\frac{1}{\tau}  \int_{0}^{\infty} dt_w \,     \frac{d}{dt_w} \left[ \int_0^{t_w} dt\, p(x,t)\right]   \int_0^{t_w} d t \, p_t(t)  \nonumber\\
&+&  \frac{1}{\tau}   \left[ \left( \int_0^{t_w} d t \, p_t(t)\right)  \left( \int_0^{t_w} dt\, p(x,t) \right) \right]_0^{\infty}
\ea
%\ba
%g(x)  &=&   -\frac{1}{\tau}  \int_{0}^{\infty} dt_p \,   p(x,t_p)   \int_0^{t_p} d t \, p_t(t)  \nonumber\\
%&+&  \frac{1}{\tau}  \left( \int_0^{\infty} d t \, p_t(t)\right)  \left( \int_0^{\infty} dt\, p(x,t) \right) 
%\ea
After some evaluation, the above expression reduces to  
\be
g(x)  =   -\frac{1}{\tau}  \int_{0}^{\infty} dt_w \,   p(x,t_w)   \int_0^{t_w} d t \, p_t(t)   +  \frac{1}{\tau}   \int_0^{\infty} dt\, p(x,t).  
\label{eq:app0b-1}
\ee
Since $\int_0^{t_w} d t \, p_t(t) =1$, we can write
$$
 \int_0^{t_w} d t \, p_t(t)  =  1 -  \int_{t_w}^{\infty} d t \, p_t(t).
$$
Inserting the above identity into Eq. (\ref{eq:app0b-1}) leads to 
\be
g(x)  =   \frac{1}{\tau}  \int_{0}^{\infty} dt_w \,   p(x,t_w)   \int_{t_w}^{\infty} d t \, p_t(t).  
\label{eq:app0b-2}
\ee
Finally, given the definition of the survival function, 
\be
S(t) = \int_t^{\infty} dt'\, p_t(t'), 
\ee
Eq. (\ref{eq:app0b-2}) recovers the expression in Eq. (\ref{eq:g}), given below 
for calrity
\be
g(x)  =   \frac{1}{\tau}  \int_{0}^{\infty} dt \,    S(t) p(x,t).  
\label{eq:app0b-3}
\ee

\section{Details on the derivation of Eq. (\ref{eq:Gpm})}
\label{sec:app0c}

In this section, we provide details of how to transform Eq. (\ref{eq:G-00}), given below for clarity, 
\be
G = \int_0^{\infty} dt \, p_t \int_{-v_0}^{v_0} dv \, p_v  \delta\left( x -  x_0  e^{-\mu K t}     
-    \frac{v\left( 1  -  e^{-\mu K t} \right)}{\mu K}  \right),
\ee
into Eq. (\ref{eq:Gpm}).  We start by rewriting the above equation as 
%\be
%G = \int_0^{\infty} dt \, p_t \int_{-v_0}^{v_0} dv \, p_v  \delta\left( x    
%-     \frac{v}{\mu K}   +  e^{-\mu K t} \left[ \frac{v}{\mu K}  -  x_0 \right]  \right).  
%\ee
\be
G = \frac{1}{\tau} \int_{-v_0}^{v_0} dv \, p_v  
 \int_0^{\infty} dt \, e^{-t/\tau} \delta\left( x    -     \frac{v}{\mu K}   +  \left( \frac{v}{\mu K}  -  x_0 \right) e^{- \mu K t}  \right), 
\label{eq:app0c-1}
\ee
where we assume an exponential distribution $p_t = \tau^{-1} e^{-t/\tau}$, and then change the sequence of 
integrations.  In the next step, we use a change of a variable
$$
s = e^{- \mu K t },  ~~~~~ ds = - \mu K s dt, 
$$
which also permits us to write $e^{- t/\tau } =  s^{\alpha}$, where $\alpha = (\tau \mu K)^{-1}$.  
Changing the variables in Eq. (\ref{eq:app0c-1}) leads to 
%\be
%G = - \alpha  \int_{-v_0}^{v_0} dv \, p_v  
% \int_0^{\infty}  ds    \, s^{\alpha-1} \delta\left( x    -     \frac{v}{\mu K}   +   \left( \frac{v}{\mu K}  -  x_0 \right) s  \right).  
%\ee
\be
G = \alpha      \int_{-v_0}^{v_0} \frac{dv}{\mu K} \,   \frac{\mu K p_v}{  |x_0 - \frac{v}{\mu K}|  }
 \int_0^{1}  ds    \, s^{\alpha-1} \delta\left( s - \frac{ x    -     \frac{v}{\mu K}}{  x_0 - \frac{v}{\mu K}  }    \right), 
\ee
where we used $\int_{-\infty}^{\infty} dx\, \delta(xa) = \frac{1}{|a|} \int_{-\infty}^{\infty} dx\, \delta(x) $.  
We next simplify the nomenclature and define 
$
y = \frac{v}{\mu K}.  
$
This allows us to write the above equation as 
\be
G = \alpha      \int_{-x_m}^{x_m} dy \,   \frac{p(y)}{  |x_0 - y|  }
 \int_0^{1}  ds    \, s^{\alpha-1} \delta\left( s - \frac{ x    -     y }{  x_0 - y }    \right),
\ee
and where $p(y)$ is related to $p_v(v)$ and is defined in Eq. (\ref{eq:py}).  
%\be
%G = \alpha      \int_{-v_0}^{v_0} dv \,   \frac{p_v}{  |x_0 - \frac{v}{\mu K}|  }
% \int_0^{1}  ds    \, s^{\alpha-1} \delta\left( s -   \left[ 1  -  \frac{ x  -  x_0 } {   \frac{v}{\mu K}  -  x_0 }  \right]     \right),
%\ee
We are now in the position to evaluate the integral over $s$.   A simple minded integration yields 
%\be
%G =  \alpha      \int_{-v_0}^{v_0} dv \,   \frac{p_v(v)}{  | x_0 - \frac{v}{\mu K} |  }   \left(   \frac{ x    -     \frac{v}{\mu K}}{  x_0 - \frac{v}{\mu K}  }  \right)^{\alpha-1}.  
%\ee
\be
G =  \alpha      \int_{-x_m}^{x_m} dy \,   \frac{p(y)}{  |x_0 - y| }   \left(   \frac{ y    -   x   } {  y -  x_0  }  \right)^{\alpha-1}, 
\ee
where 
$
x_m = \frac{v_0}{\mu K}.  
$
However, the integral over $s$ involves limits.  We can incorporates these limits indirectly 
by observing that to avoid 
imaginary numbers due to the term $ (   \frac{ y    -   x   } {  y -  x_0  }  )^{\alpha-1}$ we 
need to make sure that either $y  >  x$ and $y > x_0$, which, when implemented 
yields 
%We could leave the above equation as it is, or we could still try to get rid of the absolute value.  
%To eliminate the absolute value, we write
%\be
%G =  \alpha      \int_{-x_m}^{x} dy \,   \frac{p(y)}{  x_0 - y }   \left(   \frac{ y    -   x   } {  y -  x_0  }  \right)^{\alpha-1}, 
%\ee
%where it is assumed that $x > y$ (due to the limits) and $x_0 > y$ (so that we can ), which implies $x_0 > y$.  
$$
G_+(x,x_0)= \alpha \int_{x}^{x_m} dy\,  \frac{p(y)}{y - x_0}  \left( \frac{y - x}{y - x_0} \right)^{\alpha-1}, 
$$
or $y  <  x$ and $y < x_0$, which, when implemented yields
$$
G_-(x,x_0)= \alpha  \int_{-x_m}^{x} dy\,   \frac{p(y)}{x_0 - y}  \left( \frac{y - x}{y - x_0} \right)^{\alpha-1}.  
$$
Finally, we note that in order for both equations to be positive, $G_+(x,x_0<x)$ and $G_-(x,x_0>x)$.  We now 
recover the expressions in Eq. (\ref{eq:Gpm}).  

\section{Derivation of $\rho$ for the case $d=1$}
\label{sec:app0d}

In this section we show how the two differential equations, Eq. (\ref{eq:drho}) and Eq. (\ref{eq:dq}), 
shown again below for clarity, 
\ba
&& (x_m^2 - x^2) \rho'  =   x  \rho  -  (\alpha - 1) x_m  q, \nonumber\\ 
&& (x_m^2 - x^2) q'   =    x_m \rho  -  (\alpha - 1) x q, 
\label{eq:app0d-1}
\ea
are combined into a single differential equation.  We start by using the first equation 
above to obtain an expression for $q$, by rearrangement, and then for $q'$, by taking 
derivative of that expression.  This leads to
\ba
&& (\alpha - 1) x_m  q      =     x\rho      -    (x_m^2 - x^2) \rho'   \nonumber\\ 
&& (\alpha - 1) x_m  q'     =       \rho      +      3x \rho'    -     (x_m^2 - x^2) \rho''.  
\ea
The above equations are then inserted into the second equation in Eq. (\ref{eq:app0d-1}).  
We first substitute for $q$, which after some algebraic manipulation 
yields
$$
(x_m^2 - x^2) x_m q'   =    (x_m^2  -   x^2)\rho      +    (x_m^2 - x^2) x\rho',
$$
and then for $q'$, which after manipulation yields 
%$$
%%  (x_m^2 - x^2)\rho      +      
% -     (x_m^2 - x^2) \rho''   =    (\alpha - 2)\rho   +    (\alpha - 4) x\rho',
%$$
%These equations can be combined to yield a single equation for $\rho$, which
%is a second-order differential equation given by 
\be
0   =     (\alpha - 2) \rho      +     (\alpha - 4)  x \rho'       +      (x_m^2 - x^2)  \rho''.  
\label{eq:app0d-2}
\ee
Finally, we note that the above second-order differential equation can be written as
$$
0   =     (\alpha - 2) [x\rho]'       -    [x^2  \rho']'     +       x_m^2 \rho''.  
$$
which after integration becomes 
\be
\text{const}   =     (\alpha - 2) x\rho       -    x^2  \rho'     +       x_m^2 \rho',
\label{eq:app0d-3}
\ee
where we still need to determine the constant parameter on the left-hand-side.  
To do this, we choose the most convenient point, $x=0$.  This eliminates the first two terms on the
right-hand-side, yielding $\text{const}   =      x_m^2 \rho'(0)$.   We also know that $\rho(x)$ is an even 
function, due to symmetry of the harmonic potential.  This means that $\rho'(0) = 0$, thus, 
$\text{const}   =   0$, and Eq. (\ref{eq:app0d-3}) becomes 
\be
0   =     (\alpha - 2) x\rho        +     (x_m^2   -    x^2) \rho',
%\label{eq:diff-1d}
\ee
which recovers the first-order differential equation in Eq. (\ref{eq:diff-1d}).

%\section{Mathematica implementation of analytical approach in Sec. (\ref{sec:anal-a1})}
%\label{sec:app1}
%
%In this section we show a simple Mathematica code for generating a truncated series 
%$\rho_M(z) \propto 1 + \sum_{n=1}^M a_n z^{2n}$, for $\alpha=1$, as explained in 
%Sec. (\ref{sec:anal-a1}).  
%
%%\vspace{0.5cm}
%%
%%For $\alpha=1$
%
%\vspace{0.2cm}
%{\scriptsize 
%$M = 10;$
%
%$\text{A} = \text{Table}\left[ \frac{\text{UnitStep}[n-m-1]}{2 m- 2 n - 1} +  \frac{1}{2 m- 2 n - 1}  + 2 m \text{KroneckerDelta} [n-m], \{m,M\},\{n,M\}\right]; $
%
%$\text{b} = \text{Table}[-\frac{2}{2n - 1}, \{n, M\}];$
%
%$\text{IA} = \text{Inverse}[\text{A}];$
%
%$a = \text{IA}.\text{b};$
%
%$\text{norm} = 2 + \sum_{n=1}^{M} \frac{2 a[[n]]}{1+2n} $
%
%$\text{rho[z\_]} := \left(1 + \sum_{n=1}^{M} a[[n]] z^{2n}\right) / \text{norm}$
%}
%
%
%\section{ Arbitrary $p_t$}
%\label{sec:app2}

%\vspace{0.5cm}
%And for $\alpha=2$
%\vspace{0.2cm}
%
%{\scriptsize
%$M = 10;$
%
%$\text{A} = \text{Table}\left[ \frac{2n}{2 m - 1}  \frac{\text{UnitStep}[n-m-1]}{2 m- 2 n - 1} +  \frac{1}{2 m- 2 n - 1}  +  m \text{KroneckerDelta} [n-m], \{m,M\},\{n,M\}\right]; $
%
%$\text{b} = \text{Table}[-\frac{1}{2n - 1}, \{n, M\}];$
%
%$\text{IA} = \text{Inverse}[\text{A}];$
%
%$a = \text{IA}.\text{b};$
%
%$\text{norm} = 2 + \sum_{n=1}^{M} \frac{2 a[[n]]}{1+2n} $
%
%$\text{rho[z\_]} := \left(1 + \sum_{n=1}^{M} a[[n]]  z^{2n}\right) / \text{norm}$
%}

%------------------------------------------------
% References
%------------------------------------------------
\bibliography{RTP-harmonic}

\begin{thebibliography}{67}%
\makeatletter
\providecommand \@ifxundefined [1]{%
 \@ifx{#1\undefined}
}%
\providecommand \@ifnum [1]{%
 \ifnum #1\expandafter \@firstoftwo
 \else \expandafter \@secondoftwo
 \fi
}%
\providecommand \@ifx [1]{%
 \ifx #1\expandafter \@firstoftwo
 \else \expandafter \@secondoftwo
 \fi
}%
\providecommand \natexlab [1]{#1}%
\providecommand \enquote  [1]{``#1''}%
\providecommand \bibnamefont  [1]{#1}%
\providecommand \bibfnamefont [1]{#1}%
\providecommand \citenamefont [1]{#1}%
\providecommand \href@noop [0]{\@secondoftwo}%
\providecommand \href [0]{\begingroup \@sanitize@url \@href}%
\providecommand \@href[1]{\@@startlink{#1}\@@href}%
\providecommand \@@href[1]{\endgroup#1\@@endlink}%
\providecommand \@sanitize@url [0]{\catcode `\\12\catcode `\$12\catcode
  `\&12\catcode `\#12\catcode `\^12\catcode `\_12\catcode `\%12\relax}%
\providecommand \@@startlink[1]{}%
\providecommand \@@endlink[0]{}%
\providecommand \url  [0]{\begingroup\@sanitize@url \@url }%
\providecommand \@url [1]{\endgroup\@href {#1}{\urlprefix }}%
\providecommand \urlprefix  [0]{URL }%
\providecommand \Eprint [0]{\href }%
\providecommand \doibase [0]{http://dx.doi.org/}%
\providecommand \selectlanguage [0]{\@gobble}%
\providecommand \bibinfo  [0]{\@secondoftwo}%
\providecommand \bibfield  [0]{\@secondoftwo}%
\providecommand \translation [1]{[#1]}%
\providecommand \BibitemOpen [0]{}%
\providecommand \bibitemStop [0]{}%
\providecommand \bibitemNoStop [0]{.\EOS\space}%
\providecommand \EOS [0]{\spacefactor3000\relax}%
\providecommand \BibitemShut  [1]{\csname bibitem#1\endcsname}%
\let\auto@bib@innerbib\@empty
%</preamble>
\bibitem [{\citenamefont {Frydel}(2024{\natexlab{a}})}]{POF-Frydel-2024}%
  \BibitemOpen
  \bibfield  {author} {\bibinfo {author} {\bibfnamefont {D.}~\bibnamefont
  {Frydel}},\ }\href {\doibase 10.1063/5.0179375} {\bibfield  {journal}
  {\bibinfo  {journal} {Physics of Fluids}\ }\textbf {\bibinfo {volume} {36}},\
  \bibinfo {pages} {011910} (\bibinfo {year} {2024}{\natexlab{a}})}\BibitemShut
  {NoStop}%
\bibitem [{\citenamefont {Tailleur}\ and\ \citenamefont
  {Cates}(2008)}]{PRL-Cates-2008}%
  \BibitemOpen
  \bibfield  {author} {\bibinfo {author} {\bibfnamefont {J.}~\bibnamefont
  {Tailleur}}\ and\ \bibinfo {author} {\bibfnamefont {M.~E.}\ \bibnamefont
  {Cates}},\ }\href {\doibase 10.1103/PhysRevLett.100.218103} {\bibfield
  {journal} {\bibinfo  {journal} {Phys. Rev. Lett.}\ }\textbf {\bibinfo
  {volume} {100}},\ \bibinfo {pages} {218103} (\bibinfo {year}
  {2008})}\BibitemShut {NoStop}%
\bibitem [{\citenamefont {Tailleur}\ and\ \citenamefont
  {Cates}(2009)}]{EPL-Cates-2009}%
  \BibitemOpen
  \bibfield  {author} {\bibinfo {author} {\bibfnamefont {J.}~\bibnamefont
  {Tailleur}}\ and\ \bibinfo {author} {\bibfnamefont {M.~E.}\ \bibnamefont
  {Cates}},\ }\href {\doibase 10.1209/0295-5075/86/60002} {\bibfield  {journal}
  {\bibinfo  {journal} {Europhysics Letters}\ }\textbf {\bibinfo {volume}
  {86}},\ \bibinfo {pages} {60002} (\bibinfo {year} {2009})}\BibitemShut
  {NoStop}%
\bibitem [{\citenamefont {Grognot}\ and\ \citenamefont
  {Taute}(2021)}]{COM-Grognot-2021}%
  \BibitemOpen
  \bibfield  {author} {\bibinfo {author} {\bibfnamefont {M.}~\bibnamefont
  {Grognot}}\ and\ \bibinfo {author} {\bibfnamefont {K.~M.}\ \bibnamefont
  {Taute}},\ }\href {\doibase https://doi.org/10.1016/j.mib.2021.02.005}
  {\bibfield  {journal} {\bibinfo  {journal} {Current Opinion in Microbiology}\
  }\textbf {\bibinfo {volume} {61}},\ \bibinfo {pages} {73} (\bibinfo {year}
  {2021})}\BibitemShut {NoStop}%
\bibitem [{\citenamefont {Malakar}\ \emph {et~al.}(2018)\citenamefont
  {Malakar}, \citenamefont {Jemseena}, \citenamefont {Kundu}, \citenamefont
  {Kumar}, \citenamefont {Sabhapandit}, \citenamefont {Majumdar}, \citenamefont
  {Redner},\ and\ \citenamefont {Dhar}}]{Malakar-2018}%
  \BibitemOpen
  \bibfield  {author} {\bibinfo {author} {\bibfnamefont {K.}~\bibnamefont
  {Malakar}}, \bibinfo {author} {\bibfnamefont {V.}~\bibnamefont {Jemseena}},
  \bibinfo {author} {\bibfnamefont {A.}~\bibnamefont {Kundu}}, \bibinfo
  {author} {\bibfnamefont {K.~V.}\ \bibnamefont {Kumar}}, \bibinfo {author}
  {\bibfnamefont {S.}~\bibnamefont {Sabhapandit}}, \bibinfo {author}
  {\bibfnamefont {S.~N.}\ \bibnamefont {Majumdar}}, \bibinfo {author}
  {\bibfnamefont {S.}~\bibnamefont {Redner}}, \ and\ \bibinfo {author}
  {\bibfnamefont {A.}~\bibnamefont {Dhar}},\ }\href {\doibase
  10.1088/1742-5468/aab84f} {\bibfield  {journal} {\bibinfo  {journal} {Journal
  of Statistical Mechanics: Theory and Experiment}\ }\textbf {\bibinfo {volume}
  {2018}},\ \bibinfo {pages} {043215} (\bibinfo {year} {2018})}\BibitemShut
  {NoStop}%
\bibitem [{\citenamefont {Razin}(2020)}]{PRE-Razin-2020}%
  \BibitemOpen
  \bibfield  {author} {\bibinfo {author} {\bibfnamefont {N.}~\bibnamefont
  {Razin}},\ }\href {\doibase 10.1103/PhysRevE.102.030103} {\bibfield
  {journal} {\bibinfo  {journal} {Phys. Rev. E}\ }\textbf {\bibinfo {volume}
  {102}},\ \bibinfo {pages} {030103} (\bibinfo {year} {2020})}\BibitemShut
  {NoStop}%
\bibitem [{\citenamefont {Berg}(1983)}]{berg-1983}%
  \BibitemOpen
  \bibfield  {author} {\bibinfo {author} {\bibfnamefont {H.~C.}\ \bibnamefont
  {Berg}},\ }\href@noop {} {\emph {\bibinfo {title} {Random Walks in
  Biology}}}\ (\bibinfo  {publisher} {Princeton University Press},\ \bibinfo
  {address} {Princeton, NJ},\ \bibinfo {year} {1983})\BibitemShut {NoStop}%
\bibitem [{\citenamefont {Schnitzer}(1993)}]{PRE-Schnitzer-1993}%
  \BibitemOpen
  \bibfield  {author} {\bibinfo {author} {\bibfnamefont {M.~J.}\ \bibnamefont
  {Schnitzer}},\ }\href {\doibase 10.1103/PhysRevE.48.2553} {\bibfield
  {journal} {\bibinfo  {journal} {Phys. Rev. E}\ }\textbf {\bibinfo {volume}
  {48}},\ \bibinfo {pages} {2553} (\bibinfo {year} {1993})}\BibitemShut
  {NoStop}%
\bibitem [{\citenamefont {Romanczuk}\ \emph {et~al.}(2012)\citenamefont
  {Romanczuk}, \citenamefont {B{\"a}r}, \citenamefont {Ebeling}, \citenamefont
  {Lindner},\ and\ \citenamefont {Schimansky-Geier}}]{EPJ-Ebeling-2012}%
  \BibitemOpen
  \bibfield  {author} {\bibinfo {author} {\bibfnamefont {P.}~\bibnamefont
  {Romanczuk}}, \bibinfo {author} {\bibfnamefont {M.}~\bibnamefont {B{\"a}r}},
  \bibinfo {author} {\bibfnamefont {W.}~\bibnamefont {Ebeling}}, \bibinfo
  {author} {\bibfnamefont {B.}~\bibnamefont {Lindner}}, \ and\ \bibinfo
  {author} {\bibfnamefont {L.}~\bibnamefont {Schimansky-Geier}},\ }\href
  {\doibase 10.1140/epjst/e2012-01529-y} {\bibfield  {journal} {\bibinfo
  {journal} {The European Physical Journal Special Topics}\ }\textbf {\bibinfo
  {volume} {202}},\ \bibinfo {pages} {1} (\bibinfo {year} {2012})}\BibitemShut
  {NoStop}%
\bibitem [{\citenamefont {Fily}\ and\ \citenamefont
  {Marchetti}(2012)}]{PRL-Marchetti-2012}%
  \BibitemOpen
  \bibfield  {author} {\bibinfo {author} {\bibfnamefont {Y.}~\bibnamefont
  {Fily}}\ and\ \bibinfo {author} {\bibfnamefont {M.~C.}\ \bibnamefont
  {Marchetti}},\ }\href {\doibase 10.1103/PhysRevLett.108.235702} {\bibfield
  {journal} {\bibinfo  {journal} {Phys. Rev. Lett.}\ }\textbf {\bibinfo
  {volume} {108}},\ \bibinfo {pages} {235702} (\bibinfo {year}
  {2012})}\BibitemShut {NoStop}%
\bibitem [{\citenamefont {Marchetti}\ \emph {et~al.}(2013)\citenamefont
  {Marchetti}, \citenamefont {Joanny}, \citenamefont {Ramaswamy}, \citenamefont
  {Liverpool}, \citenamefont {Prost}, \citenamefont {Rao},\ and\ \citenamefont
  {Simha}}]{RMP-Marchetti-2013}%
  \BibitemOpen
  \bibfield  {author} {\bibinfo {author} {\bibfnamefont {M.~C.}\ \bibnamefont
  {Marchetti}}, \bibinfo {author} {\bibfnamefont {J.~F.}\ \bibnamefont
  {Joanny}}, \bibinfo {author} {\bibfnamefont {S.}~\bibnamefont {Ramaswamy}},
  \bibinfo {author} {\bibfnamefont {T.~B.}\ \bibnamefont {Liverpool}}, \bibinfo
  {author} {\bibfnamefont {J.}~\bibnamefont {Prost}}, \bibinfo {author}
  {\bibfnamefont {M.}~\bibnamefont {Rao}}, \ and\ \bibinfo {author}
  {\bibfnamefont {R.~A.}\ \bibnamefont {Simha}},\ }\href {\doibase
  10.1103/RevModPhys.85.1143} {\bibfield  {journal} {\bibinfo  {journal} {Rev.
  Mod. Phys.}\ }\textbf {\bibinfo {volume} {85}},\ \bibinfo {pages} {1143}
  (\bibinfo {year} {2013})}\BibitemShut {NoStop}%
\bibitem [{\citenamefont {Maggi}\ \emph {et~al.}(2015)\citenamefont {Maggi},
  \citenamefont {Marconi}, \citenamefont {Gnan},\ and\ \citenamefont
  {Di~Leonardo}}]{SR-Maggi-2015}%
  \BibitemOpen
  \bibfield  {author} {\bibinfo {author} {\bibfnamefont {C.}~\bibnamefont
  {Maggi}}, \bibinfo {author} {\bibfnamefont {U.~M.~B.}\ \bibnamefont
  {Marconi}}, \bibinfo {author} {\bibfnamefont {N.}~\bibnamefont {Gnan}}, \
  and\ \bibinfo {author} {\bibfnamefont {R.}~\bibnamefont {Di~Leonardo}},\
  }\href {\doibase 10.1038/srep10742} {\bibfield  {journal} {\bibinfo
  {journal} {Scientific Reports}\ }\textbf {\bibinfo {volume} {5}},\ \bibinfo
  {pages} {10742} (\bibinfo {year} {2015})}\BibitemShut {NoStop}%
\bibitem [{\citenamefont {Solon}\ \emph
  {et~al.}(2015{\natexlab{a}})\citenamefont {Solon}, \citenamefont {Cates},\
  and\ \citenamefont {Tailleur}}]{EPJ-Cates-2015}%
  \BibitemOpen
  \bibfield  {author} {\bibinfo {author} {\bibfnamefont {A.~P.}\ \bibnamefont
  {Solon}}, \bibinfo {author} {\bibfnamefont {M.~E.}\ \bibnamefont {Cates}}, \
  and\ \bibinfo {author} {\bibfnamefont {J.}~\bibnamefont {Tailleur}},\ }\href
  {\doibase 10.1140/epjst/e2015-02457-0} {\bibfield  {journal} {\bibinfo
  {journal} {The European Physical Journal Special Topics}\ }\textbf {\bibinfo
  {volume} {224}},\ \bibinfo {pages} {1231} (\bibinfo {year}
  {2015}{\natexlab{a}})}\BibitemShut {NoStop}%
\bibitem [{\citenamefont {Étienne Fodor}\ and\ \citenamefont
  {Cristina Marchetti}(2018)}]{PhysA-Fodor-2018}%
  \BibitemOpen
  \bibfield  {author} {\bibinfo {author} {\bibnamefont {Étienne Fodor}}\ and\
  \bibinfo {author} {\bibfnamefont {M.}~\bibnamefont {Cristina Marchetti}},\
  }\href {\doibase https://doi.org/10.1016/j.physa.2017.12.137} {\bibfield
  {journal} {\bibinfo  {journal} {Physica A: Statistical Mechanics and its
  Applications}\ }\textbf {\bibinfo {volume} {504}},\ \bibinfo {pages} {106}
  (\bibinfo {year} {2018})},\ \bibinfo {note} {lecture Notes of the 14th
  International Summer School on Fundamental Problems in Statistical
  Physics}\BibitemShut {NoStop}%
\bibitem [{\citenamefont {Shankar}\ and\ \citenamefont
  {Marchetti}(2018)}]{PRE-Shankar-2018}%
  \BibitemOpen
  \bibfield  {author} {\bibinfo {author} {\bibfnamefont {S.}~\bibnamefont
  {Shankar}}\ and\ \bibinfo {author} {\bibfnamefont {M.~C.}\ \bibnamefont
  {Marchetti}},\ }\href {\doibase 10.1103/PhysRevE.98.020604} {\bibfield
  {journal} {\bibinfo  {journal} {Phys. Rev. E}\ }\textbf {\bibinfo {volume}
  {98}},\ \bibinfo {pages} {020604} (\bibinfo {year} {2018})}\BibitemShut
  {NoStop}%
\bibitem [{\citenamefont {Das}\ \emph {et~al.}(2018)\citenamefont {Das},
  \citenamefont {Gompper},\ and\ \citenamefont {Winkler}}]{NJP-Gompper-2018}%
  \BibitemOpen
  \bibfield  {author} {\bibinfo {author} {\bibfnamefont {S.}~\bibnamefont
  {Das}}, \bibinfo {author} {\bibfnamefont {G.}~\bibnamefont {Gompper}}, \ and\
  \bibinfo {author} {\bibfnamefont {R.~G.}\ \bibnamefont {Winkler}},\ }\href
  {\doibase 10.1088/1367-2630/aa9d4b} {\bibfield  {journal} {\bibinfo
  {journal} {New Journal of Physics}\ }\textbf {\bibinfo {volume} {20}},\
  \bibinfo {pages} {015001} (\bibinfo {year} {2018})}\BibitemShut {NoStop}%
\bibitem [{\citenamefont {Caraglio}\ and\ \citenamefont
  {Franosch}(2022)}]{PRL-Caraglio-2022}%
  \BibitemOpen
  \bibfield  {author} {\bibinfo {author} {\bibfnamefont {M.}~\bibnamefont
  {Caraglio}}\ and\ \bibinfo {author} {\bibfnamefont {T.}~\bibnamefont
  {Franosch}},\ }\href {\doibase 10.1103/PhysRevLett.129.158001} {\bibfield
  {journal} {\bibinfo  {journal} {Phys. Rev. Lett.}\ }\textbf {\bibinfo
  {volume} {129}},\ \bibinfo {pages} {158001} (\bibinfo {year}
  {2022})}\BibitemShut {NoStop}%
\bibitem [{\citenamefont {Nakul}\ and\ \citenamefont
  {Gopalakrishnan}(2023)}]{PRE-Manoj-2023}%
  \BibitemOpen
  \bibfield  {author} {\bibinfo {author} {\bibfnamefont {U.}~\bibnamefont
  {Nakul}}\ and\ \bibinfo {author} {\bibfnamefont {M.}~\bibnamefont
  {Gopalakrishnan}},\ }\href {\doibase 10.1103/PhysRevE.108.024121} {\bibfield
  {journal} {\bibinfo  {journal} {Phys. Rev. E}\ }\textbf {\bibinfo {volume}
  {108}},\ \bibinfo {pages} {024121} (\bibinfo {year} {2023})}\BibitemShut
  {NoStop}%
\bibitem [{\citenamefont {Caporusso}\ \emph {et~al.}(2023)\citenamefont
  {Caporusso}, \citenamefont {Cugliandolo}, \citenamefont {Digregorio},
  \citenamefont {Gonnella}, \citenamefont {Levis},\ and\ \citenamefont
  {Suma}}]{PRL-Suma-2023b}%
  \BibitemOpen
  \bibfield  {author} {\bibinfo {author} {\bibfnamefont {C.~B.}\ \bibnamefont
  {Caporusso}}, \bibinfo {author} {\bibfnamefont {L.~F.}\ \bibnamefont
  {Cugliandolo}}, \bibinfo {author} {\bibfnamefont {P.}~\bibnamefont
  {Digregorio}}, \bibinfo {author} {\bibfnamefont {G.}~\bibnamefont
  {Gonnella}}, \bibinfo {author} {\bibfnamefont {D.}~\bibnamefont {Levis}}, \
  and\ \bibinfo {author} {\bibfnamefont {A.}~\bibnamefont {Suma}},\ }\href
  {\doibase 10.1103/PhysRevLett.131.068201} {\bibfield  {journal} {\bibinfo
  {journal} {Phys. Rev. Lett.}\ }\textbf {\bibinfo {volume} {131}},\ \bibinfo
  {pages} {068201} (\bibinfo {year} {2023})}\BibitemShut {NoStop}%
\bibitem [{\citenamefont {Semeraro}\ \emph {et~al.}(2024)\citenamefont
  {Semeraro}, \citenamefont {Negro}, \citenamefont {Suma}, \citenamefont
  {Corberi},\ and\ \citenamefont {Gonnella}}]{EPL-Suma-2024}%
  \BibitemOpen
  \bibfield  {author} {\bibinfo {author} {\bibfnamefont {M.}~\bibnamefont
  {Semeraro}}, \bibinfo {author} {\bibfnamefont {G.}~\bibnamefont {Negro}},
  \bibinfo {author} {\bibfnamefont {A.}~\bibnamefont {Suma}}, \bibinfo {author}
  {\bibfnamefont {F.}~\bibnamefont {Corberi}}, \ and\ \bibinfo {author}
  {\bibfnamefont {G.}~\bibnamefont {Gonnella}},\ }\href {\doibase
  10.1209/0295-5075/ad895e} {\bibfield  {journal} {\bibinfo  {journal}
  {Europhysics Letters}\ }\textbf {\bibinfo {volume} {148}},\ \bibinfo {pages}
  {37001} (\bibinfo {year} {2024})}\BibitemShut {NoStop}%
\bibitem [{\citenamefont {Caporusso}\ \emph {et~al.}(2024)\citenamefont
  {Caporusso}, \citenamefont {Cugliandolo}, \citenamefont {Digregorio},
  \citenamefont {Gonnella},\ and\ \citenamefont {Suma}}]{SF-Suma-2024}%
  \BibitemOpen
  \bibfield  {author} {\bibinfo {author} {\bibfnamefont {C.~B.}\ \bibnamefont
  {Caporusso}}, \bibinfo {author} {\bibfnamefont {L.~F.}\ \bibnamefont
  {Cugliandolo}}, \bibinfo {author} {\bibfnamefont {P.}~\bibnamefont
  {Digregorio}}, \bibinfo {author} {\bibfnamefont {G.}~\bibnamefont
  {Gonnella}}, \ and\ \bibinfo {author} {\bibfnamefont {A.}~\bibnamefont
  {Suma}},\ }\href {\doibase 10.1039/D4SM00200H} {\bibfield  {journal}
  {\bibinfo  {journal} {Soft Matter}\ }\textbf {\bibinfo {volume} {20}},\
  \bibinfo {pages} {4208} (\bibinfo {year} {2024})}\BibitemShut {NoStop}%
\bibitem [{\citenamefont {Archer}\ and\ \citenamefont
  {Ebbens}()}]{AdvSci-2023}%
  \BibitemOpen
  \bibfield  {author} {\bibinfo {author} {\bibfnamefont {R.~J.}\ \bibnamefont
  {Archer}}\ and\ \bibinfo {author} {\bibfnamefont {S.~J.}\ \bibnamefont
  {Ebbens}},\ }\href {\doibase https://doi.org/10.1002/advs.202303154}
  {\bibfield  {journal} {\bibinfo  {journal} {Advanced Science}\ }\textbf
  {\bibinfo {volume} {10}},\ \bibinfo {pages} {2303154}}\BibitemShut {NoStop}%
\bibitem [{\citenamefont {Szamel}(2014)}]{PRE-Szamel-2014}%
  \BibitemOpen
  \bibfield  {author} {\bibinfo {author} {\bibfnamefont {G.}~\bibnamefont
  {Szamel}},\ }\href {\doibase 10.1103/PhysRevE.90.012111} {\bibfield
  {journal} {\bibinfo  {journal} {Phys. Rev. E}\ }\textbf {\bibinfo {volume}
  {90}},\ \bibinfo {pages} {012111} (\bibinfo {year} {2014})}\BibitemShut
  {NoStop}%
\bibitem [{\citenamefont {Martin}\ \emph {et~al.}(2021)\citenamefont {Martin},
  \citenamefont {O'Byrne}, \citenamefont {Cates}, \citenamefont {Fodor},
  \citenamefont {Nardini}, \citenamefont {Tailleur},\ and\ \citenamefont {van
  Wijland}}]{PRE-Fodor-2021}%
  \BibitemOpen
  \bibfield  {author} {\bibinfo {author} {\bibfnamefont {D.}~\bibnamefont
  {Martin}}, \bibinfo {author} {\bibfnamefont {J.}~\bibnamefont {O'Byrne}},
  \bibinfo {author} {\bibfnamefont {M.~E.}\ \bibnamefont {Cates}}, \bibinfo
  {author} {\bibfnamefont {E.}~\bibnamefont {Fodor}}, \bibinfo {author}
  {\bibfnamefont {C.}~\bibnamefont {Nardini}}, \bibinfo {author} {\bibfnamefont
  {J.}~\bibnamefont {Tailleur}}, \ and\ \bibinfo {author} {\bibfnamefont
  {F.}~\bibnamefont {van Wijland}},\ }\href {\doibase
  10.1103/PhysRevE.103.032607} {\bibfield  {journal} {\bibinfo  {journal}
  {Phys. Rev. E}\ }\textbf {\bibinfo {volume} {103}},\ \bibinfo {pages}
  {032607} (\bibinfo {year} {2021})}\BibitemShut {NoStop}%
\bibitem [{\citenamefont {Caprini}\ \emph {et~al.}(2019)\citenamefont
  {Caprini}, \citenamefont {Marconi}, \citenamefont {Puglisi},\ and\
  \citenamefont {Vulpiani}}]{JSTAT-Caprini-2019}%
  \BibitemOpen
  \bibfield  {author} {\bibinfo {author} {\bibfnamefont {L.}~\bibnamefont
  {Caprini}}, \bibinfo {author} {\bibfnamefont {U.~M.~B.}\ \bibnamefont
  {Marconi}}, \bibinfo {author} {\bibfnamefont {A.}~\bibnamefont {Puglisi}}, \
  and\ \bibinfo {author} {\bibfnamefont {A.}~\bibnamefont {Vulpiani}},\ }\href
  {\doibase 10.1088/1742-5468/ab14dd} {\bibfield  {journal} {\bibinfo
  {journal} {Journal of Statistical Mechanics: Theory and Experiment}\ }\textbf
  {\bibinfo {volume} {2019}},\ \bibinfo {pages} {053203} (\bibinfo {year}
  {2019})}\BibitemShut {NoStop}%
\bibitem [{\citenamefont {Semeraro}\ \emph {et~al.}(2023)\citenamefont
  {Semeraro}, \citenamefont {Gonnella}, \citenamefont {Suma},\ and\
  \citenamefont {Zamparo}}]{PRL-Suma-2023}%
  \BibitemOpen
  \bibfield  {author} {\bibinfo {author} {\bibfnamefont {M.}~\bibnamefont
  {Semeraro}}, \bibinfo {author} {\bibfnamefont {G.}~\bibnamefont {Gonnella}},
  \bibinfo {author} {\bibfnamefont {A.}~\bibnamefont {Suma}}, \ and\ \bibinfo
  {author} {\bibfnamefont {M.}~\bibnamefont {Zamparo}},\ }\href {\doibase
  10.1103/PhysRevLett.131.158302} {\bibfield  {journal} {\bibinfo  {journal}
  {Phys. Rev. Lett.}\ }\textbf {\bibinfo {volume} {131}},\ \bibinfo {pages}
  {158302} (\bibinfo {year} {2023})}\BibitemShut {NoStop}%
\bibitem [{\citenamefont {Pal}\ and\ \citenamefont
  {Sabhapandit}(2013)}]{PRE-Pal-2013}%
  \BibitemOpen
  \bibfield  {author} {\bibinfo {author} {\bibfnamefont {A.}~\bibnamefont
  {Pal}}\ and\ \bibinfo {author} {\bibfnamefont {S.}~\bibnamefont
  {Sabhapandit}},\ }\href {\doibase 10.1103/PhysRevE.87.022138} {\bibfield
  {journal} {\bibinfo  {journal} {Phys. Rev. E}\ }\textbf {\bibinfo {volume}
  {87}},\ \bibinfo {pages} {022138} (\bibinfo {year} {2013})}\BibitemShut
  {NoStop}%
\bibitem [{\citenamefont {Santra}\ \emph {et~al.}(2021)\citenamefont {Santra},
  \citenamefont {Das},\ and\ \citenamefont {Nath}}]{JOP-Santra-2021}%
  \BibitemOpen
  \bibfield  {author} {\bibinfo {author} {\bibfnamefont {I.}~\bibnamefont
  {Santra}}, \bibinfo {author} {\bibfnamefont {S.}~\bibnamefont {Das}}, \ and\
  \bibinfo {author} {\bibfnamefont {S.~K.}\ \bibnamefont {Nath}},\ }\href
  {\doibase 10.1088/1751-8121/ac12a0} {\bibfield  {journal} {\bibinfo
  {journal} {Journal of Physics A: Mathematical and Theoretical}\ }\textbf
  {\bibinfo {volume} {54}},\ \bibinfo {pages} {334001} (\bibinfo {year}
  {2021})}\BibitemShut {NoStop}%
\bibitem [{\citenamefont {Gupta}\ \emph {et~al.}(2020)\citenamefont {Gupta},
  \citenamefont {Plata}, \citenamefont {Kundu},\ and\ \citenamefont
  {Pal}}]{JOPA-Gupta-2020}%
  \BibitemOpen
  \bibfield  {author} {\bibinfo {author} {\bibfnamefont {D.}~\bibnamefont
  {Gupta}}, \bibinfo {author} {\bibfnamefont {C.~A.}\ \bibnamefont {Plata}},
  \bibinfo {author} {\bibfnamefont {A.}~\bibnamefont {Kundu}}, \ and\ \bibinfo
  {author} {\bibfnamefont {A.}~\bibnamefont {Pal}},\ }\href {\doibase
  10.1088/1751-8121/abcf0b} {\bibfield  {journal} {\bibinfo  {journal} {Journal
  of Physics A: Mathematical and Theoretical}\ }\textbf {\bibinfo {volume}
  {54}},\ \bibinfo {pages} {025003} (\bibinfo {year} {2020})}\BibitemShut
  {NoStop}%
\bibitem [{\citenamefont {Gupta}\ \emph {et~al.}(2021)\citenamefont {Gupta},
  \citenamefont {Pal},\ and\ \citenamefont {Kundu}}]{JSTAT-Gupta-2021}%
  \BibitemOpen
  \bibfield  {author} {\bibinfo {author} {\bibfnamefont {D.}~\bibnamefont
  {Gupta}}, \bibinfo {author} {\bibfnamefont {A.}~\bibnamefont {Pal}}, \ and\
  \bibinfo {author} {\bibfnamefont {A.}~\bibnamefont {Kundu}},\ }\href
  {\doibase 10.1088/1742-5468/abefdf} {\bibfield  {journal} {\bibinfo
  {journal} {Journal of Statistical Mechanics: Theory and Experiment}\ }\textbf
  {\bibinfo {volume} {2021}},\ \bibinfo {pages} {043202} (\bibinfo {year}
  {2021})}\BibitemShut {NoStop}%
\bibitem [{\citenamefont {Alston}\ \emph {et~al.}(2022)\citenamefont {Alston},
  \citenamefont {Cocconi},\ and\ \citenamefont {Bertrand}}]{JOPA-Alston-2022}%
  \BibitemOpen
  \bibfield  {author} {\bibinfo {author} {\bibfnamefont {H.}~\bibnamefont
  {Alston}}, \bibinfo {author} {\bibfnamefont {L.}~\bibnamefont {Cocconi}}, \
  and\ \bibinfo {author} {\bibfnamefont {T.}~\bibnamefont {Bertrand}},\ }\href
  {\doibase 10.1088/1751-8121/ac726b} {\bibfield  {journal} {\bibinfo
  {journal} {Journal of Physics A: Mathematical and Theoretical}\ }\textbf
  {\bibinfo {volume} {55}},\ \bibinfo {pages} {274004} (\bibinfo {year}
  {2022})}\BibitemShut {NoStop}%
\bibitem [{\citenamefont {Biroli}\ \emph {et~al.}(2024)\citenamefont {Biroli},
  \citenamefont {Kulkarni}, \citenamefont {Majumdar},\ and\ \citenamefont
  {Schehr}}]{PRE-Schehr-2024}%
  \BibitemOpen
  \bibfield  {author} {\bibinfo {author} {\bibfnamefont {M.}~\bibnamefont
  {Biroli}}, \bibinfo {author} {\bibfnamefont {M.}~\bibnamefont {Kulkarni}},
  \bibinfo {author} {\bibfnamefont {S.~N.}\ \bibnamefont {Majumdar}}, \ and\
  \bibinfo {author} {\bibfnamefont {G.}~\bibnamefont {Schehr}},\ }\href
  {\doibase 10.1103/PhysRevE.109.L032106} {\bibfield  {journal} {\bibinfo
  {journal} {Phys. Rev. E}\ }\textbf {\bibinfo {volume} {109}},\ \bibinfo
  {pages} {L032106} (\bibinfo {year} {2024})}\BibitemShut {NoStop}%
\bibitem [{\citenamefont {Frydel}(2024{\natexlab{b}})}]{PRE-Frydel-2024}%
  \BibitemOpen
  \bibfield  {author} {\bibinfo {author} {\bibfnamefont {D.}~\bibnamefont
  {Frydel}},\ }\href {\doibase 10.1103/PhysRevE.110.024613} {\bibfield
  {journal} {\bibinfo  {journal} {Phys. Rev. E}\ }\textbf {\bibinfo {volume}
  {110}},\ \bibinfo {pages} {024613} (\bibinfo {year}
  {2024}{\natexlab{b}})}\BibitemShut {NoStop}%
\bibitem [{\citenamefont {Frydel}(2022{\natexlab{a}})}]{PRE-Fyrdel-2022}%
  \BibitemOpen
  \bibfield  {author} {\bibinfo {author} {\bibfnamefont {D.}~\bibnamefont
  {Frydel}},\ }\href {\doibase 10.1103/PhysRevE.105.034113} {\bibfield
  {journal} {\bibinfo  {journal} {Phys. Rev. E}\ }\textbf {\bibinfo {volume}
  {105}},\ \bibinfo {pages} {034113} (\bibinfo {year}
  {2022}{\natexlab{a}})}\BibitemShut {NoStop}%
\bibitem [{\citenamefont {Padmanabha}\ \emph {et~al.}(2023)\citenamefont
  {Padmanabha}, \citenamefont {Busiello}, \citenamefont {Maritan},\ and\
  \citenamefont {Gupta}}]{PRE-Gupta-2023}%
  \BibitemOpen
  \bibfield  {author} {\bibinfo {author} {\bibfnamefont {P.}~\bibnamefont
  {Padmanabha}}, \bibinfo {author} {\bibfnamefont {D.~M.}\ \bibnamefont
  {Busiello}}, \bibinfo {author} {\bibfnamefont {A.}~\bibnamefont {Maritan}}, \
  and\ \bibinfo {author} {\bibfnamefont {D.}~\bibnamefont {Gupta}},\ }\href
  {\doibase 10.1103/PhysRevE.107.014129} {\bibfield  {journal} {\bibinfo
  {journal} {Phys. Rev. E}\ }\textbf {\bibinfo {volume} {107}},\ \bibinfo
  {pages} {014129} (\bibinfo {year} {2023})}\BibitemShut {NoStop}%
\bibitem [{\citenamefont {Angelani}\ \emph {et~al.}(2014)\citenamefont
  {Angelani}, \citenamefont {Di~Leonardo},\ and\ \citenamefont
  {Paoluzzi}}]{EPJE-Angelani-2014}%
  \BibitemOpen
  \bibfield  {author} {\bibinfo {author} {\bibfnamefont {L.}~\bibnamefont
  {Angelani}}, \bibinfo {author} {\bibfnamefont {R.}~\bibnamefont
  {Di~Leonardo}}, \ and\ \bibinfo {author} {\bibfnamefont {M.}~\bibnamefont
  {Paoluzzi}},\ }\href {\doibase 10.1140/epje/i2014-14059-4} {\bibfield
  {journal} {\bibinfo  {journal} {The European Physical Journal E}\ }\textbf
  {\bibinfo {volume} {37}},\ \bibinfo {pages} {59} (\bibinfo {year}
  {2014})}\BibitemShut {NoStop}%
\bibitem [{\citenamefont {Mori}\ \emph {et~al.}(2020)\citenamefont {Mori},
  \citenamefont {Le~Doussal}, \citenamefont {Majumdar},\ and\ \citenamefont
  {Schehr}}]{PRL-Schehr-2020}%
  \BibitemOpen
  \bibfield  {author} {\bibinfo {author} {\bibfnamefont {F.}~\bibnamefont
  {Mori}}, \bibinfo {author} {\bibfnamefont {P.}~\bibnamefont {Le~Doussal}},
  \bibinfo {author} {\bibfnamefont {S.~N.}\ \bibnamefont {Majumdar}}, \ and\
  \bibinfo {author} {\bibfnamefont {G.}~\bibnamefont {Schehr}},\ }\href
  {\doibase 10.1103/PhysRevLett.124.090603} {\bibfield  {journal} {\bibinfo
  {journal} {Phys. Rev. Lett.}\ }\textbf {\bibinfo {volume} {124}},\ \bibinfo
  {pages} {090603} (\bibinfo {year} {2020})}\BibitemShut {NoStop}%
\bibitem [{\citenamefont {Bruyne}\ \emph {et~al.}(2021)\citenamefont {Bruyne},
  \citenamefont {Majumdar},\ and\ \citenamefont {Schehr}}]{JSTAT-Scher-2021}%
  \BibitemOpen
  \bibfield  {author} {\bibinfo {author} {\bibfnamefont {B.~D.}\ \bibnamefont
  {Bruyne}}, \bibinfo {author} {\bibfnamefont {S.~N.}\ \bibnamefont
  {Majumdar}}, \ and\ \bibinfo {author} {\bibfnamefont {G.}~\bibnamefont
  {Schehr}},\ }\href {\doibase 10.1088/1742-5468/abf5d5} {\bibfield  {journal}
  {\bibinfo  {journal} {Journal of Statistical Mechanics: Theory and
  Experiment}\ }\textbf {\bibinfo {volume} {2021}},\ \bibinfo {pages} {043211}
  (\bibinfo {year} {2021})}\BibitemShut {NoStop}%
\bibitem [{\citenamefont {Singh}\ and\ \citenamefont
  {Kundu}(2021)}]{PRE-Kundu-2021}%
  \BibitemOpen
  \bibfield  {author} {\bibinfo {author} {\bibfnamefont {P.}~\bibnamefont
  {Singh}}\ and\ \bibinfo {author} {\bibfnamefont {A.}~\bibnamefont {Kundu}},\
  }\href {\doibase 10.1103/PhysRevE.103.042119} {\bibfield  {journal} {\bibinfo
   {journal} {Phys. Rev. E}\ }\textbf {\bibinfo {volume} {103}},\ \bibinfo
  {pages} {042119} (\bibinfo {year} {2021})}\BibitemShut {NoStop}%
\bibitem [{\citenamefont {Basu}\ \emph {et~al.}(2020)\citenamefont {Basu},
  \citenamefont {Majumdar}, \citenamefont {Rosso}, \citenamefont
  {Sabhapandit},\ and\ \citenamefont {Schehr}}]{JPA-Basu-2020}%
  \BibitemOpen
  \bibfield  {author} {\bibinfo {author} {\bibfnamefont {U.}~\bibnamefont
  {Basu}}, \bibinfo {author} {\bibfnamefont {S.~N.}\ \bibnamefont {Majumdar}},
  \bibinfo {author} {\bibfnamefont {A.}~\bibnamefont {Rosso}}, \bibinfo
  {author} {\bibfnamefont {S.}~\bibnamefont {Sabhapandit}}, \ and\ \bibinfo
  {author} {\bibfnamefont {G.}~\bibnamefont {Schehr}},\ }\href {\doibase
  10.1088/1751-8121/ab6af0} {\bibfield  {journal} {\bibinfo  {journal} {Journal
  of Physics A: Mathematical and Theoretical}\ }\textbf {\bibinfo {volume}
  {53}},\ \bibinfo {pages} {09LT01} (\bibinfo {year} {2020})}\BibitemShut
  {NoStop}%
\bibitem [{\citenamefont {Frydel}(2021{\natexlab{a}})}]{JSTAT-Frydel-2021}%
  \BibitemOpen
  \bibfield  {author} {\bibinfo {author} {\bibfnamefont {D.}~\bibnamefont
  {Frydel}},\ }\href {\doibase 10.1088/1742-5468/ac1665} {\bibfield  {journal}
  {\bibinfo  {journal} {Journal of Statistical Mechanics: Theory and
  Experiment}\ }\textbf {\bibinfo {volume} {2021}},\ \bibinfo {pages} {083220}
  (\bibinfo {year} {2021}{\natexlab{a}})}\BibitemShut {NoStop}%
\bibitem [{\citenamefont {Frydel}(2022{\natexlab{b}})}]{POF-Frydel-2022}%
  \BibitemOpen
  \bibfield  {author} {\bibinfo {author} {\bibfnamefont {D.}~\bibnamefont
  {Frydel}},\ }\href {\doibase 10.1063/5.0080058} {\bibfield  {journal}
  {\bibinfo  {journal} {Physics of Fluids}\ }\textbf {\bibinfo {volume} {34}},\
  \bibinfo {pages} {027111} (\bibinfo {year} {2022}{\natexlab{b}})}\BibitemShut
  {NoStop}%
\bibitem [{\citenamefont {Breoni}\ \emph {et~al.}(2022)\citenamefont {Breoni},
  \citenamefont {Schwarzendahl}, \citenamefont {Blossey},\ and\ \citenamefont
  {L{\"o}wen}}]{EPJE-Hartmut-2022}%
  \BibitemOpen
  \bibfield  {author} {\bibinfo {author} {\bibfnamefont {D.}~\bibnamefont
  {Breoni}}, \bibinfo {author} {\bibfnamefont {F.~J.}\ \bibnamefont
  {Schwarzendahl}}, \bibinfo {author} {\bibfnamefont {R.}~\bibnamefont
  {Blossey}}, \ and\ \bibinfo {author} {\bibfnamefont {H.}~\bibnamefont
  {L{\"o}wen}},\ }\href {\doibase 10.1140/epje/s10189-022-00238-7} {\bibfield
  {journal} {\bibinfo  {journal} {The European Physical Journal E}\ }\textbf
  {\bibinfo {volume} {45}},\ \bibinfo {pages} {83} (\bibinfo {year}
  {2022})}\BibitemShut {NoStop}%
\bibitem [{\citenamefont {Frydel}(2022{\natexlab{c}})}]{PRE-106-Frydel-2022}%
  \BibitemOpen
  \bibfield  {author} {\bibinfo {author} {\bibfnamefont {D.}~\bibnamefont
  {Frydel}},\ }\href {\doibase 10.1103/PhysRevE.106.024121} {\bibfield
  {journal} {\bibinfo  {journal} {Phys. Rev. E}\ }\textbf {\bibinfo {volume}
  {106}},\ \bibinfo {pages} {024121} (\bibinfo {year}
  {2022}{\natexlab{c}})}\BibitemShut {NoStop}%
\bibitem [{\citenamefont {Smith}\ \emph {et~al.}(2022)\citenamefont {Smith},
  \citenamefont {Le~Doussal}, \citenamefont {Majumdar},\ and\ \citenamefont
  {Schehr}}]{PRE-Scher-2022}%
  \BibitemOpen
  \bibfield  {author} {\bibinfo {author} {\bibfnamefont {N.~R.}\ \bibnamefont
  {Smith}}, \bibinfo {author} {\bibfnamefont {P.}~\bibnamefont {Le~Doussal}},
  \bibinfo {author} {\bibfnamefont {S.~N.}\ \bibnamefont {Majumdar}}, \ and\
  \bibinfo {author} {\bibfnamefont {G.}~\bibnamefont {Schehr}},\ }\href
  {\doibase 10.1103/PhysRevE.106.054133} {\bibfield  {journal} {\bibinfo
  {journal} {Phys. Rev. E}\ }\textbf {\bibinfo {volume} {106}},\ \bibinfo
  {pages} {054133} (\bibinfo {year} {2022})}\BibitemShut {NoStop}%
\bibitem [{\citenamefont {Smith}\ and\ \citenamefont
  {Farago}(2022)}]{Smith-PRE-2022a}%
  \BibitemOpen
  \bibfield  {author} {\bibinfo {author} {\bibfnamefont {N.~R.}\ \bibnamefont
  {Smith}}\ and\ \bibinfo {author} {\bibfnamefont {O.}~\bibnamefont {Farago}},\
  }\href {\doibase 10.1103/PhysRevE.106.054118} {\bibfield  {journal} {\bibinfo
   {journal} {Phys. Rev. E}\ }\textbf {\bibinfo {volume} {106}},\ \bibinfo
  {pages} {054118} (\bibinfo {year} {2022})}\BibitemShut {NoStop}%
\bibitem [{\citenamefont {Frydel}(2023)}]{POF-Frydel-2023}%
  \BibitemOpen
  \bibfield  {author} {\bibinfo {author} {\bibfnamefont {D.}~\bibnamefont
  {Frydel}},\ }\href {\doibase 10.1063/5.0173374} {\bibfield  {journal}
  {\bibinfo  {journal} {Physics of Fluids}\ }\textbf {\bibinfo {volume} {35}},\
  \bibinfo {pages} {101905} (\bibinfo {year} {2023})}\BibitemShut {NoStop}%
\bibitem [{\citenamefont {Dhar}\ \emph {et~al.}(2019)\citenamefont {Dhar},
  \citenamefont {Kundu}, \citenamefont {Majumdar}, \citenamefont
  {Sabhapandit},\ and\ \citenamefont {Schehr}}]{PRE-Dhar-2019}%
  \BibitemOpen
  \bibfield  {author} {\bibinfo {author} {\bibfnamefont {A.}~\bibnamefont
  {Dhar}}, \bibinfo {author} {\bibfnamefont {A.}~\bibnamefont {Kundu}},
  \bibinfo {author} {\bibfnamefont {S.~N.}\ \bibnamefont {Majumdar}}, \bibinfo
  {author} {\bibfnamefont {S.}~\bibnamefont {Sabhapandit}}, \ and\ \bibinfo
  {author} {\bibfnamefont {G.}~\bibnamefont {Schehr}},\ }\href {\doibase
  10.1103/PhysRevE.99.032132} {\bibfield  {journal} {\bibinfo  {journal} {Phys.
  Rev. E}\ }\textbf {\bibinfo {volume} {99}},\ \bibinfo {pages} {032132}
  (\bibinfo {year} {2019})}\BibitemShut {NoStop}%
\bibitem [{\citenamefont {Farago}\ and\ \citenamefont
  {Smith}(2024)}]{PRE-Farago-2024}%
  \BibitemOpen
  \bibfield  {author} {\bibinfo {author} {\bibfnamefont {O.}~\bibnamefont
  {Farago}}\ and\ \bibinfo {author} {\bibfnamefont {N.~R.}\ \bibnamefont
  {Smith}},\ }\href {\doibase 10.1103/PhysRevE.109.044121} {\bibfield
  {journal} {\bibinfo  {journal} {Phys. Rev. E}\ }\textbf {\bibinfo {volume}
  {109}},\ \bibinfo {pages} {044121} (\bibinfo {year} {2024})}\BibitemShut
  {NoStop}%
\bibitem [{\citenamefont {Roberts}\ and\ \citenamefont
  {Zhen}(2023)}]{PRE-connor-2023}%
  \BibitemOpen
  \bibfield  {author} {\bibinfo {author} {\bibfnamefont {C.}~\bibnamefont
  {Roberts}}\ and\ \bibinfo {author} {\bibfnamefont {Z.}~\bibnamefont {Zhen}},\
  }\href {\doibase 10.1103/PhysRevE.108.014139} {\bibfield  {journal} {\bibinfo
   {journal} {Phys. Rev. E}\ }\textbf {\bibinfo {volume} {108}},\ \bibinfo
  {pages} {014139} (\bibinfo {year} {2023})}\BibitemShut {NoStop}%
\bibitem [{\citenamefont {Doussal}\ \emph {et~al.}(2020)\citenamefont
  {Doussal}, \citenamefont {Majumdar},\ and\ \citenamefont
  {Schehr}}]{EPL-Doussal-2020}%
  \BibitemOpen
  \bibfield  {author} {\bibinfo {author} {\bibfnamefont {P.~L.}\ \bibnamefont
  {Doussal}}, \bibinfo {author} {\bibfnamefont {S.~N.}\ \bibnamefont
  {Majumdar}}, \ and\ \bibinfo {author} {\bibfnamefont {G.}~\bibnamefont
  {Schehr}},\ }\href {\doibase 10.1209/0295-5075/130/40002} {\bibfield
  {journal} {\bibinfo  {journal} {Europhysics Letters}\ }\textbf {\bibinfo
  {volume} {130}},\ \bibinfo {pages} {40002} (\bibinfo {year}
  {2020})}\BibitemShut {NoStop}%
\bibitem [{\citenamefont {de~Pirey}\ and\ \citenamefont {van
  Wijland}(2023)}]{JSTAT-Wijland-2023}%
  \BibitemOpen
  \bibfield  {author} {\bibinfo {author} {\bibfnamefont {T.~A.}\ \bibnamefont
  {de~Pirey}}\ and\ \bibinfo {author} {\bibfnamefont {F.}~\bibnamefont {van
  Wijland}},\ }\href {\doibase 10.1088/1742-5468/ace42d} {\bibfield  {journal}
  {\bibinfo  {journal} {Journal of Statistical Mechanics: Theory and
  Experiment}\ }\textbf {\bibinfo {volume} {2023}},\ \bibinfo {pages} {093202}
  (\bibinfo {year} {2023})}\BibitemShut {NoStop}%
\bibitem [{\citenamefont {Angelani}(2015)}]{JPA-Angelani-2015}%
  \BibitemOpen
  \bibfield  {author} {\bibinfo {author} {\bibfnamefont {L.}~\bibnamefont
  {Angelani}},\ }\href {\doibase 10.1088/1751-8113/48/49/495003} {\bibfield
  {journal} {\bibinfo  {journal} {Journal of Physics A: Mathematical and
  Theoretical}\ }\textbf {\bibinfo {volume} {48}},\ \bibinfo {pages} {495003}
  (\bibinfo {year} {2015})}\BibitemShut {NoStop}%
\bibitem [{\citenamefont {Bressloff}(2023)}]{JSTAT-Bressloff-2023}%
  \BibitemOpen
  \bibfield  {author} {\bibinfo {author} {\bibfnamefont {P.~C.}\ \bibnamefont
  {Bressloff}},\ }\href {\doibase 10.1088/1742-5468/accce2} {\bibfield
  {journal} {\bibinfo  {journal} {Journal of Statistical Mechanics: Theory and
  Experiment}\ }\textbf {\bibinfo {volume} {2023}},\ \bibinfo {pages} {043208}
  (\bibinfo {year} {2023})}\BibitemShut {NoStop}%
\bibitem [{\citenamefont {Woillez}\ \emph {et~al.}(2019)\citenamefont
  {Woillez}, \citenamefont {Zhao}, \citenamefont {Kafri}, \citenamefont
  {Lecomte},\ and\ \citenamefont {Tailleur}}]{PRL-Tailleur-2019}%
  \BibitemOpen
  \bibfield  {author} {\bibinfo {author} {\bibfnamefont {E.}~\bibnamefont
  {Woillez}}, \bibinfo {author} {\bibfnamefont {Y.}~\bibnamefont {Zhao}},
  \bibinfo {author} {\bibfnamefont {Y.}~\bibnamefont {Kafri}}, \bibinfo
  {author} {\bibfnamefont {V.}~\bibnamefont {Lecomte}}, \ and\ \bibinfo
  {author} {\bibfnamefont {J.}~\bibnamefont {Tailleur}},\ }\href {\doibase
  10.1103/PhysRevLett.122.258001} {\bibfield  {journal} {\bibinfo  {journal}
  {Phys. Rev. Lett.}\ }\textbf {\bibinfo {volume} {122}},\ \bibinfo {pages}
  {258001} (\bibinfo {year} {2019})}\BibitemShut {NoStop}%
\bibitem [{\citenamefont {Detcheverry}(2015)}]{EPL-Detcheverry-2015}%
  \BibitemOpen
  \bibfield  {author} {\bibinfo {author} {\bibfnamefont {F.}~\bibnamefont
  {Detcheverry}},\ }\href {\doibase 10.1209/0295-5075/111/60002} {\bibfield
  {journal} {\bibinfo  {journal} {Europhysics Letters}\ }\textbf {\bibinfo
  {volume} {111}},\ \bibinfo {pages} {60002} (\bibinfo {year}
  {2015})}\BibitemShut {NoStop}%
\bibitem [{\citenamefont {Dean}\ \emph {et~al.}(2021)\citenamefont {Dean},
  \citenamefont {Majumdar},\ and\ \citenamefont {Schawe}}]{PRE-Dean-2021}%
  \BibitemOpen
  \bibfield  {author} {\bibinfo {author} {\bibfnamefont {D.~S.}\ \bibnamefont
  {Dean}}, \bibinfo {author} {\bibfnamefont {S.~N.}\ \bibnamefont {Majumdar}},
  \ and\ \bibinfo {author} {\bibfnamefont {H.}~\bibnamefont {Schawe}},\ }\href
  {\doibase 10.1103/PhysRevE.103.012130} {\bibfield  {journal} {\bibinfo
  {journal} {Phys. Rev. E}\ }\textbf {\bibinfo {volume} {103}},\ \bibinfo
  {pages} {012130} (\bibinfo {year} {2021})}\BibitemShut {NoStop}%
\bibitem [{\citenamefont {Angelani}(2017)}]{JPA-Angelani-2017}%
  \BibitemOpen
  \bibfield  {author} {\bibinfo {author} {\bibfnamefont {L.}~\bibnamefont
  {Angelani}},\ }\href {\doibase 10.1088/1751-8121/aa734c} {\bibfield
  {journal} {\bibinfo  {journal} {Journal of Physics A: Mathematical and
  Theoretical}\ }\textbf {\bibinfo {volume} {50}},\ \bibinfo {pages} {325601}
  (\bibinfo {year} {2017})}\BibitemShut {NoStop}%
\bibitem [{\citenamefont {Klinger}\ \emph {et~al.}(2022)\citenamefont
  {Klinger}, \citenamefont {Voituriez},\ and\ \citenamefont
  {B\'enichou}}]{PRL-Klinger-2022}%
  \BibitemOpen
  \bibfield  {author} {\bibinfo {author} {\bibfnamefont {J.}~\bibnamefont
  {Klinger}}, \bibinfo {author} {\bibfnamefont {R.}~\bibnamefont {Voituriez}},
  \ and\ \bibinfo {author} {\bibfnamefont {O.}~\bibnamefont {B\'enichou}},\
  }\href {\doibase 10.1103/PhysRevLett.129.140603} {\bibfield  {journal}
  {\bibinfo  {journal} {Phys. Rev. Lett.}\ }\textbf {\bibinfo {volume} {129}},\
  \bibinfo {pages} {140603} (\bibinfo {year} {2022})}\BibitemShut {NoStop}%
\bibitem [{\citenamefont {Dutta}\ \emph {et~al.}(2025)\citenamefont {Dutta},
  \citenamefont {Kundu},\ and\ \citenamefont {Basu}}]{Chaos-Basu-2025}%
  \BibitemOpen
  \bibfield  {author} {\bibinfo {author} {\bibfnamefont {D.}~\bibnamefont
  {Dutta}}, \bibinfo {author} {\bibfnamefont {A.}~\bibnamefont {Kundu}}, \ and\
  \bibinfo {author} {\bibfnamefont {U.}~\bibnamefont {Basu}},\ }\href {\doibase
  10.1063/5.0250965} {\bibfield  {journal} {\bibinfo  {journal} {Chaos: An
  Interdisciplinary Journal of Nonlinear Science}\ }\textbf {\bibinfo {volume}
  {35}},\ \bibinfo {pages} {033109} (\bibinfo {year} {2025})},\ \Eprint
  {http://arxiv.org/abs/https://pubs.aip.org/aip/cha/article-pdf/doi/10.1063/5.0250965/20420036/033109\_1\_5.0250965.pdf}
  {https://pubs.aip.org/aip/cha/article-pdf/doi/10.1063/5.0250965/20420036/033109\_1\_5.0250965.pdf}
  \BibitemShut {NoStop}%
\bibitem [{\citenamefont {Rotter}\ and\ \citenamefont
  {Gigan}(2017)}]{RevModPhys-Rotter-2017}%
  \BibitemOpen
  \bibfield  {author} {\bibinfo {author} {\bibfnamefont {S.}~\bibnamefont
  {Rotter}}\ and\ \bibinfo {author} {\bibfnamefont {S.}~\bibnamefont {Gigan}},\
  }\href {\doibase 10.1103/RevModPhys.89.015005} {\bibfield  {journal}
  {\bibinfo  {journal} {Rev. Mod. Phys.}\ }\textbf {\bibinfo {volume} {89}},\
  \bibinfo {pages} {015005} (\bibinfo {year} {2017})}\BibitemShut {NoStop}%
\bibitem [{\citenamefont {Solon}\ \emph
  {et~al.}(2015{\natexlab{b}})\citenamefont {Solon}, \citenamefont {Fily},
  \citenamefont {Baskaran}, \citenamefont {Cates}, \citenamefont {Kafri},
  \citenamefont {Kardar},\ and\ \citenamefont {Tailleur}}]{pressure-2015}%
  \BibitemOpen
  \bibfield  {author} {\bibinfo {author} {\bibfnamefont {A.~P.}\ \bibnamefont
  {Solon}}, \bibinfo {author} {\bibfnamefont {Y.}~\bibnamefont {Fily}},
  \bibinfo {author} {\bibfnamefont {A.}~\bibnamefont {Baskaran}}, \bibinfo
  {author} {\bibfnamefont {M.~E.}\ \bibnamefont {Cates}}, \bibinfo {author}
  {\bibfnamefont {Y.}~\bibnamefont {Kafri}}, \bibinfo {author} {\bibfnamefont
  {M.}~\bibnamefont {Kardar}}, \ and\ \bibinfo {author} {\bibfnamefont
  {J.}~\bibnamefont {Tailleur}},\ }\href {\doibase 10.1038/nphys3377}
  {\bibfield  {journal} {\bibinfo  {journal} {Nature Physics}\ }\textbf
  {\bibinfo {volume} {11}},\ \bibinfo {pages} {673} (\bibinfo {year}
  {2015}{\natexlab{b}})}\BibitemShut {NoStop}%
\bibitem [{\citenamefont {Meyn}\ \emph {et~al.}(2009)\citenamefont {Meyn},
  \citenamefont {Tweedie},\ and\ \citenamefont {Glynn}}]{Glynn-2009}%
  \BibitemOpen
  \bibfield  {author} {\bibinfo {author} {\bibfnamefont {S.}~\bibnamefont
  {Meyn}}, \bibinfo {author} {\bibfnamefont {R.~L.}\ \bibnamefont {Tweedie}}, \
  and\ \bibinfo {author} {\bibfnamefont {P.~W.}\ \bibnamefont {Glynn}},\
  }\href@noop {} {\emph {\bibinfo {title} {Markov Chains and Stochastic
  Stability}}},\ \bibinfo {edition} {2nd}\ ed.,\ Cambridge Mathematical
  Library\ (\bibinfo  {publisher} {Cambridge University Press},\ \bibinfo
  {year} {2009})\BibitemShut {NoStop}%
\bibitem [{\citenamefont {Nartallo-Kaluarachchi}\ \emph
  {et~al.}(2024)\citenamefont {Nartallo-Kaluarachchi}, \citenamefont {Asllani},
  \citenamefont {Deco}, \citenamefont {Kringelbach}, \citenamefont {Goriely},\
  and\ \citenamefont {Lambiotte}}]{PRE-Ramon-2024}%
  \BibitemOpen
  \bibfield  {author} {\bibinfo {author} {\bibfnamefont {R.}~\bibnamefont
  {Nartallo-Kaluarachchi}}, \bibinfo {author} {\bibfnamefont {M.}~\bibnamefont
  {Asllani}}, \bibinfo {author} {\bibfnamefont {G.}~\bibnamefont {Deco}},
  \bibinfo {author} {\bibfnamefont {M.~L.}\ \bibnamefont {Kringelbach}},
  \bibinfo {author} {\bibfnamefont {A.}~\bibnamefont {Goriely}}, \ and\
  \bibinfo {author} {\bibfnamefont {R.}~\bibnamefont {Lambiotte}},\ }\href
  {\doibase 10.1103/PhysRevE.110.034313} {\bibfield  {journal} {\bibinfo
  {journal} {Phys. Rev. E}\ }\textbf {\bibinfo {volume} {110}},\ \bibinfo
  {pages} {034313} (\bibinfo {year} {2024})}\BibitemShut {NoStop}%
\bibitem [{\citenamefont {Frydel}(2021{\natexlab{b}})}]{PRE-103-Frydel-2021}%
  \BibitemOpen
  \bibfield  {author} {\bibinfo {author} {\bibfnamefont {D.}~\bibnamefont
  {Frydel}},\ }\href {\doibase 10.1103/PhysRevE.103.052603} {\bibfield
  {journal} {\bibinfo  {journal} {Phys. Rev. E}\ }\textbf {\bibinfo {volume}
  {103}},\ \bibinfo {pages} {052603} (\bibinfo {year}
  {2021}{\natexlab{b}})}\BibitemShut {NoStop}%
\bibitem [{\citenamefont {Forrester}(2010)}]{Forrester}%
  \BibitemOpen
  \bibfield  {author} {\bibinfo {author} {\bibfnamefont {P.~J.}\ \bibnamefont
  {Forrester}},\ }\href {https://cds.cern.ch/record/1315169} {\emph {\bibinfo
  {title} {{Log‑Gases and Random Matrices}}}},\ London Mathematical Society
  monographs\ (\bibinfo  {publisher} {Princeton University Press},\ \bibinfo
  {address} {Princeton, NJ},\ \bibinfo {year} {2010})\BibitemShut {NoStop}%
\bibitem [{\citenamefont {Detcheverry}(2017)}]{PRE-Fran-2017}%
  \BibitemOpen
  \bibfield  {author} {\bibinfo {author} {\bibfnamefont {F.}~\bibnamefont
  {Detcheverry}},\ }\href {\doibase 10.1103/PhysRevE.96.012415} {\bibfield
  {journal} {\bibinfo  {journal} {Phys. Rev. E}\ }\textbf {\bibinfo {volume}
  {96}},\ \bibinfo {pages} {012415} (\bibinfo {year} {2017})}\BibitemShut
  {NoStop}%
\end{thebibliography}%

%------------------------------------------------

\end{document}